\documentclass[12pt]{iopart}
\usepackage{graphicx} %,amsmath}

\newcommand {\la} {\langle}\newcommand {\ra} {\rangle}
\newcommand {\beq} {\begin{eqnarray}}
\newcommand {\eeqn} [1] {\label{#1} \end{eqnarray}}
\newcommand {\eol} {\nonumber \\}
\newcommand {\ve} [1] {\mbox{\boldmath $#1$}}
\newcommand {\dem} {\mbox{$\frac{1}{2}$}}

\begin{document}

\title{Modelling overlap functions for one-nucleon removal: role of the
effective three-nucleon force }
\author{ N. K. Timofeyuk }
\address{Department of Physics, Faculty of Engineering and Physical Sciences,\\ University of Surrey, Guildford, GU2 7XH, UK}
\ead{N.Timofeyuk@surrey.ac.uk}
%\date{August 2018}

\begin{abstract}
One-nucleon overlap functions, needed for nucleon-removal reaction calculations, are solutions of an inhomogeneous equation with the source term defined by the wave functions of the initial and final nuclear states and interaction between the removed nucleon with the rest. 
The source term approach (STA) allows  the overlaps with correct asymptotic decrease to be modelled while using nuclear many-body functions calculated in minimal model spaces. By properly choosing the removed nucleon interaction the minimum-model-space STA can reproduce reduced values of spectroscopic factors extracted from nucleon-removal reactions and predicts isospin asymmetry in the spectroscopic factor reduction.
 It is well-known that model space truncation leads to the appearance of  higher-order induced forces, with three-nucleon force being the most important. In this paper the role of such a force on the source term calculation is studied. Applications to one-nucleon removal from double-magic nuclei show that three-nucleon force improves the description of available phenomenological overlap functions and reduces  isospin asymmetry in spectroscopic factors. 
\end{abstract}

\section{Introduction}

The theoretical description of  transfer, breakup, knockout and nucleon capture reactions,
in which one nucleon is added to or removed from  the target or projectile,
  requires knowledge of one-nucleon overlap functions. %are an important input for calculations of nuclear reactions , such as transfer, breakup, knockout and nucleon capture. 
 The overlap function is defined as
\beq
I(\ve{r}) = \sqrt{A} \, \la \psi_{A-1}| \psi_A\ra = \sqrt{A} \int d\{\xi\}_{A-1} \psi_{A-1}^*(\{\xi\}_{A-1}) \psi_A(\{\xi\}_A),
\eeqn{OF}
where the wave functions $\psi_A$ and $\psi_{A-1}$ of nuclei $A$ and $A-1$ depend on internal Jacobi coordinates $\{\xi\}_A$ and $\{\xi\}_{A-1}$ for $A$ and $A-1$ nucleons, respectively, while $\ve{r}$ is the last Jacobi coordinate describing the position of the removed nucleon with respect to the centre of mass of $A-1$.
The definition of the overlap integral often includes the factor $\sqrt{A}$ to avoid multiplication of the cross section by the factor of $A$ that arises due to the antisymmetrization (see discussion in   \cite{TJPPNP}). This factor arises in the isospin formalism that treats neutron and proton as two different projections of one particle. Note that if neutrons and protons are treated separately then $\sqrt{N_A}$ or $\sqrt{Z_A}$ is used instead of $\sqrt{A}$ for neutron or proton removal, respectively, where $N_A$ ($Z_A$) is the number of neutrons (protons) in $A$.

The calculations of $I(\ve{r})$ performed so far within modern microscopic nuclear models and  the challenges these models face are reviewed in \cite{TJPPNP,Tim14}. Perhaps the main problem with these models, except for the lightest nuclei with  $A \leq 7$, is that they cannot reproduce the correct asymptotic decrease of $I(\ve{r})$ at large $r$ which is so important for nucleon removal reactions. For some light nuclei ab-initio calculations give the correct asymptotic behaviour when binary channels are explicitly included in the total wave function and some not well-defined parameters of the nucleon-nucleon (NN) and three-nucleon (3N) forces are tuned to get nucleon separation energy  in the channel of interest to be in agreement with the experimental value. Examples of such calculations can be found in \cite{Nav11} and \cite{Cal16} for $^8$B and $^{11}$Be, respectively.

%In a recent ab-initio calculations of the $\la ^{10}{\rm Be} |^{11}{\rm Be} \ra $ overlap, where the latest version of the nucleon-nucleon (NN) and three-nucleon (3N) potentials from the chiral effective field theory from  was used, the correct asymptotics has been achieved by adjusting the cut-off parameter in the 3N regularization, which is precisely defined. What about 8B?

It is possible to construct the function $I(\ve{r})$ with guaranteed correct asymptotic decrease if instead of direct evaluation of the integral (\ref{OF}) the function $I(\ve{r})$ is found from the inhomogeneous equation \cite{Pin65,Phi68,Ban85}
\beq
(T_N  + S_N) I (\ve{r}) = -\la \psi_{A-1} |\sum_{i \in A-1} V_{iA} | \psi_A\ra,
\eeqn{ineq}
where $T_N$ is the kinetic energy operator associated with variable $\ve{r}$ and $S_N $ is the (positive) nucleon separation energy. Because of the short range of the two-body NN  interaction $V_{ij}$    the right hand side of Eq. (\ref{ineq}) goes to zero at large $r$ inducing an exponential decrease in $I(\ve{r})$. For proton removal, a point-charge Coulomb potential should be added to both sides of equation (\ref{ineq}) to cancel the long-range contributions from the Coulomb NN force in the source term $\la \psi_{A-1} |\sum\limits_{i \in A-1} V_{ij} | \psi_A\ra$ \cite{Tim98}. Using an experimental value of $S_N$ in Eq. (\ref{ineq}) makes its solutions $I(\ve{r})$  applicable for nucleon removal reaction studies provided the  nuclear models for $A-1$ and $A$ 
describe  $\psi_{A-1}$ and $\psi_A$ in the internal nuclear region reasonably well.

Early work that used inhomogeneous equation (\ref{ineq}) can be found in \cite{Pin65,Phi68,Ban85} and in references therein. In a more recent approach of \cite{Tim98,Tim09,Tim10,Tim11,Tim13} 
the source term was calculated using 0$\hbar \omega$ harmonic oscillator wave functions  of $\psi_{A-1}$ and $\psi_A$  for $0p$-shell and $A \geq 16$  double-closed-shell nuclei. The advantage of the harmonic oscillator wave functions is that they allow  the centre-of-mass motion to be removed exactly, which is very important for being consistent with reaction theories which require the overlaps to be a function of $\ve{r}$. The wrong tails of the oscillator wave functions do not create a  problem because of the short range of the nuclear interaction $V_{ij}$ that suppress contributions from the asymptotic parts of $\psi_{A-1}$ and $\psi_A$. 

Truncating the model space to $0\hbar\omega$ results in the need to use effective interactions $V_{ij}$ in the source term calculations, which may be different from effective interactions used to generate $\psi_{A-1}$ and $\psi_A$ \cite{Tim09,Tim10}. The choice of the effective NN interaction made in \cite{Tim98,Tim09,Tim10,Tim11,Tim13}  resulted in a reduced spectroscopic strength of nuclear states of stable nuclei from $^4$He to $^{208}$Pb in a uniform manner and in some asymmetry in this reduction for removing weakly- and strongly-bound nucleons. This asymmetry remains a hot topic in modern nuclear physics research \cite{Gom18,Vaq20}.

It is known from many-body calculations that truncating model spaces leads  to the appearance of many-body interactions dominated by an induced 3N force \cite{Fel98,Fur12}. The contribution of this force could potentially be larger than that of the bare 3N force. One can expect that the induced 3N force could play an important role in the source term calculations as well. All source-term calculations performed until now used the NN interaction only. The aim of this paper is to investigate the role of an effective 3N force for generating the overlap functions from the inhomogeneous equation (\ref{ineq}). A simple model of the 3N force is used with parameters  that provide a reasonable agreement between the overlaps calculated for double-closed shell nuclei and those extracted from $(e,e'p)$ reactions. The paper first reviews the properties of the  overlap functions in Section 2, then describes the source-term formalism with 3N forces in Section 3. The basic input and its uncertainties are discussed in Section 4. Numerical calculations with 3N force are presented Section 5 and conclusions are drawn in Section 6. Useful expressions for 3N matrix elements needed to evaluate the source term are given in the Appendix.

%Such a truncation of the model space requires effective two-body interaction and empirically it was found that the M3Y from xxx does a good job.

\section{Overlap functions and their properties}

The overlap function (\ref{OF}) depends on angular momenta $J_{A-1}$ and $J_A$ and their projections $M_{A-1}$ and $M_A$ of nuclei $A-1$ and $A$, respectively, and should carry these indices as well,
 $I^{J_{A-1}J_A}_{M_{A-1}M_A}(\ve{r})$. Reaction codes that use the overlaps as an input are based on the partial wave decomposition  
\beq
\fl 
I^{J_{A-1}J_A}_{M_{A-1}M_A}(\ve{r}) = \sum_{lj m_lm_j \sigma } (lm_l \dem \sigma | j m_j) (j m_j J_{A-1} M_{A-1} | J_A M_A) I_{lj}^{J_{A-1}J_A}(r) Y_{lm_l}(\hat{\ve{r}}) \phi_{\sigma} \phi_{\tau},  
\eeqn{OFexp}
where $I_{lj}^{J_{A-1}J_A}(r)$ is the radial part, $Y_{lm_l}(\hat{\ve{r}})$  represents spherical function with the orbital momentum $l$ and its projection $m_l$, $\phi_{\sigma}$ is the spin function of the nucleon in $A$ not belonging to $A-1$ and the Clebsch-Gordan coefficients couple all the angular momenta.  Isospin formalism is used in this paper, which means that the nucleon isospin wave function  $\phi_{\tau}$  is also present in the expansion (\ref{OFexp}). Strictly speaking, in this case the overlap function should carry indices denoted by isospins $T_{A-1}$ and $T_A$ of nuclei $A-1$ and $A$ as well  their projections $M_{T_{A-1}}$ and $M_{T_A}$, respectively, and the Clebsch-Gordan coefficient $( T_{A-1} M_{T_{A-1}} \dem \tau | T_A M_{T_A})$ responsible for coupling these isospins should be present  in (\ref{OFexp}). This  paper assumes that the isospin Clebsch-Gordon coefficient is included in $I_{lj}^{J_{A-1}J_A}(r)$ while isospin variables are omitted. Normally, this does not cause any confusion given that most nuclear states are isospin-pure.

The norm of the radial part of the overlap function is called the spectroscopic factor,
\beq
S_{lj}^{J_{A-1}J_A} = \int_0^{\infty} dr\, r^2 \left[I_{lj}^{J_{A-1}J_A}(r)\right]^2.
\eeqn{SFdef}
Usually, the cross sections of one-nucleon removal reactions correlate with the spectroscopic factors, suggesting that such reactions are an excellent tool for extracting them from experiments. The interest in such activity has been triggered by the  shell model interpretation of spectroscopic factors  $S_{l j}^{J_{A-1}J_A}$  proposed in \cite{Mac60} where they were shown to be represented as reduced matrix elements of particle creation operators and interpreted as a measure of the occupancy  of nucleon orbits $lj$ in $A-1$ and $A$. In reality, even in their simplest versions, one-nucleon removal amplitudes do not contains spectroscopic factors, being  convolutions of $I_{lj}^{J_{A-1}J_A}(r)$ with other quantities such as distorted waves. Quite often, it is only the surface and/or external parts of $I_{lj}^{J_{A-1}J_A}(r)$ that contribute to the reaction amplitude while $S_{lj}^{J_{A-1}J_A}$ is determined mainly by contributions from $I_{lj}^{J_{A-1}J_A}(r)$ at small $r$ in Eq. (\ref{SFdef}). This means that the experimental determination of  $S_{lj}^{J_{A-1}J_A}$ from surface-dominated reactions has inherent uncertainties (see discussions in \cite{TJPPNP,Tim14} and references therein).

The inhomogeneous equation (\ref{ineq}) with the source term that vanishes at large $r$ dictates the well-known asymptotic behaviour of the overlap function $I_{lj}^{J_{A-1}J_A}(r)$ at $r \rightarrow \infty$,
\beq
I_{lj}^{J_{A-1}J_A} (r) \approx C_{l j}^{J_{A-1}J_A} W_{-\eta,l+1/2}(2\kappa r)/r,
\eeqn{ANCdef}
where $W$ is the Whittaker function, $\kappa = \sqrt{2 \mu S_N}/\hbar$, %$S_N$ is the separation energy of a nucleon from $A$ 
and $C_{l j}^{J_{A-1}J_A}$ is the asymptotic normalization coefficient (ANC). For peripheral one-nucleon removal reactions the    cross sections can be therefore factorized in terms of ANC squared and the latter can be extracted from these reactions with a significantly better accuracy than the spectroscopic factors. Eq. (\ref{ineq}) gives opportunity to calculate the ANCs and then compare them to those obtained from experiment.

This paper presents calculations of $I_{lj} (r)$, $S_{lj}$ and $C_{lj} $ (nuclear spins are omitted for brevity) and also gives the quantities $b _{lj} = C _{lj}/\sqrt{S_{lj}}$  called single-particle ANCs. They determine the magnitude of the tail of the single-particle wave function $\varphi_{l j}(r)$ which is often used to model the overlap function as $I_{lj}(r) = \sqrt{S_{lj}} \,\varphi_{l j}(r)$ in the nucleon-removal reaction studies. In general, $b_{lj}^2$  is correlated with the root-mean square radius $\la r^2 \ra_{lj}^{1/2}$ of this overlap, also calculated in this paper and defined as
\beq
\la r^2 \ra_{lj} = \frac{ \int_0^{\infty} dr\, r^4  I_{lj}^2(r)}
{\int_0^{\infty} dr\, r^2  I_{lj}^2(r) }.
\eeqn{rmsdef}
For overlap functions associated with the removal of nucleons with small separation energies the $\la r^2 \ra_{lj}^{1/2}$  is often associated with the radius of the halo state.

\section{Source term with three-nucleon interaction}

It is easy to show that when 3N interactions are included the source term will have an additional contribution 
\beq
S^{(3N)}(\ve{r}) = \la \psi_{A-1} |\sum_{i<j \in A-1} W_{ijA} | \psi_A\ra.
\eeqn{3NST}
to the r.h.s. of (\ref{ineq}). The source term corresponding to 0$\hbar\omega$  harmonic oscillator wave functions $\psi_{A-1}$ and $\psi_A$ and two-body NN interactions has been extensively studied in \cite{Tim10,Tim11,Tim13} where analytical expressions for basic two-nucleon matrix elements are given. An important feature of these calculations was treating $\psi_{A-1}$ and $\psi_A$  as translation-invariant. %Effects of centre-of-mass were investigated in \cite{Tim11} where it was shown that its effect on spectroscopic factors and ANCs is larger than the $1/A$.
The role played by the centre-of-mass was investigated in \cite{Tim11} where it was shown that its effect on spectroscopic factors and ANCs results in a percentage correction that is larger than the factor $1/A$.

Here we extend these methods to calculate $S^{(3N)}(\ve{r})$ assuming that either $A-1$ or $A$ is a double-magic nucleus.  Using harmonic oscillator wave functions allows   the source term (\ref{3NST}) to be related to a more convenient matrix element that includes Slater determinants $\Phi_{A-1}(\ve{r}_1,\ve{r}_2,...,\ve{r}_{A-1})$ and $\Phi_{A}(\ve{r}_1,\ve{r}_2,...,\ve{r}_{A})$ describing $A-1$ and $A$ in an arbitrary coordinate system in individual nucleon coordinates $\ve{r}_i$. The generalization of the formalism developed in \cite{Tim11}, sections II and III, gives
 \beq
 \fl
{S}^{(3N)}(\ve{r}) = \frac{\mu^{3/2}}{(2\pi)^3} \left(\frac{A-1}{A}\right)^{3/4}e^{ \frac{\mu}{2A}\frac{r^2}{2b^2}}
\int d\ve{k}\,e^{-i\mu \ve{k}\ve{r}}
 e^{\varepsilon k^2b^2}
\la \Phi_{A-1} e^{i  \ve{k}\ve{r}_A}|
   \sum_{i<j \in A-1} W_{ijA} | \Phi_{A} \ra,
   \eol
\eeqn{SourceTerm}
where $\mu = \frac{2A-2}{2A-1}$, $\varepsilon = \frac{1}{4A-2}$ and $b$ is the oscillator length. The integration in the matrix element in (\ref{SourceTerm}) that includes $\Phi_{A-1}$ and $\Phi_A$ is carried out over all individual coordinates $\ve{r}_1, \ve{r}_2 , ..., \ve{r}_A$. Since the nuclear wave functions are antisymmetric and the 3N force ${W}$ is symmetric with respect to nucleon permutations then this matrix element is
\beq
\fl
\la \Phi_{A-1} e^{i  \ve{k}\ve{r}_A} |
\sum_{i<j \in A-1}{W}_{ijA}
|\Phi_A\ra
 = \frac{(A-1)A}{2} \,
 \la \Phi_{A-1} 
e^{i  \ve{k}\ve{r}_A} |{W}_{A-2,{A-1},A}
|\Phi_A\ra.
\eeqn{ME}
Since the model wave functions $\Phi_{A-1}$ and $\Phi_A$ are represented by single-particle wave functions $\varphi_{\alpha_i}(\ve{r}_i)$, where $\alpha$ denotes a set of quantum numbers $\{n,l,j,m_j,\tau_j\}$, the calculation of matrix element (\ref{ME}) reduces to a calculation of the basic matrix elements
\beq
\la \varphi_{\alpha'_1}(\ve{r}_1)
\varphi_{\alpha'_2}(\ve{r}_2) e^{i  \ve{k}\ve{r}_3}\mid %W(\ve{r}_1,\ve{r}_2,\ve{r}_3) 
W_{123}\mid
\varphi_{\alpha_1}(\ve{r}_1)\varphi_{\alpha_2}(\ve{r}_2)
\varphi_{\alpha_3}(\ve{r}_3) \ra.
\eeqn{indME}
In this paper the   3N potential of the following structure will be used:
\beq
%W(\ve{r}_1,\ve{r}_2,\ve{r}_3) = W_0(\ve{r}_1,\ve{r}_2,\ve{r}_3)
%\eol
\fl W_{123} = W^{(0)}_{123 }
+
\sum_{i\neq k<j \neq k}\left[(\ve{\sigma}_i \cdot \ve{\sigma}_j) W_{k,ij}^{(\sigma)} 
 +
(\ve{\tau}_i \cdot \ve{\tau}_j) W_{k,ij}^{(\tau)} 
+
(\ve{\sigma}_i \cdot \ve{\sigma}_j)(\ve{\tau}_i \cdot \ve{\tau}_j) W_{k,ij}^{(\sigma\tau)}\right].
\eeqn{3Nmodel}
The radial dependence of the spacial parts $W_{i,jk}^{(\cal O)}$ is assumed for simplicity to be a function of the hyperradius $\rho_{ijk} = \sqrt{(r_{ij}^2 + r_{jk}^2 + r_{ik}^2)/3}$ of the three-nucleon system $ijk$ and to have a gaussian form:
\beq
W_{ijk} (\rho_{ijk}) = {W}_0  \exp \left(-\frac{1}{3}\frac{r_{ij}^2 + r_{jk}^2 + r_{ik}^2}{\rho^2_0}\right).
\eeqn{WHC}
Both $W_0$  and $\rho_0$ should carry additional indices associated with a particular choice of the operator ${\cal O}$ which are not shown here.
Analytical expressions for the matrix element (\ref{indME}) of such a force are given in the Appendix.

\section{Input to source term  calculations}

In this paper  numerical results were obtained using the same two-body  NN force as in previous publications \cite{Tim09,Tim10,Tim11,Tim13}. It is one of M3Y interactions from \cite{M3Y}, called   M3YE here, that fits harmonic oscillator matrix elements extracted from the NN phase shifts in \cite{Elliot}. The dependence of the STA results on the NN potential choice has been extensively discussed in \cite{Tim98,Tim09,Tim10}. For proton removal, the source term also contains a Coulomb correction, defined by the expression
\beq
S^{({\rm Coul})}(\ve{r}) = \la \psi_{A-1} |\sum_{i \in A-1
} V^{\rm Coul}_{iA} - \frac{  Z_{A-1} e^2}{r} | \psi_A\ra,
\eeqn{CoulombST}
which in all previous publications has been directly added to the nuclear part of the source term in all calculations. However, it should be noticed that  expression (\ref{CoulombST}) can be split into two terms with the second one being  just an overlap integral itself times the Coulomb interaction of the proton with the $A-1$ nucleus. Calculating this overlap directly within a restricted model space in most cases gives a different normalization (or spectroscopic factor) to the one obtained within the STA thus introducing some inconsistency to the method. In this paper, a new approach has been adopted in which the Coulomb correction (\ref{CoulombST}) was multiplied  by $\sqrt{S_{\rm STA}/S_{\rm SM}}$, where $S_{\rm STA}$ is the spectroscopic factor obtained from the norm of the STA overlap function while $S_{\rm SM}$ is the standard spectroscopic factor associated with the shell model wave functions $\psi_{A-1}$ and $\psi_A$ used. The introduction of such a factor guarantees consistency between the overlap normalisation used in the Coulomb correction calculations and the final result for the  overlap function $I(\ve{r})$. However,  this factor should be applied to the matrix element of $\sum\limits_{i \in A-1} V^{\rm Coul}_{iA}$ as well in order to cut  off the long-range contributions   from  the Coulomb tails. Physically, this would mean with working within small model spaces should result in introduction of the effective interactions for the Coulomb potentials as well and that such interactions could be obtained by a simple renormalisation of the Coulomb force. 

Fig.1 shows a comparison between two different treatments of the Coulomb correction for $s_{1/2}$ proton removal  from $^{40}$Ca and $^{208}$Pb. One can see that the Coulomb correction mainly affects the overlap integral in the internal region only increasing  significantly with the nuclear charge.  The spectroscopic factors  are reduced by 5$\%$ when the Coulomb corrections are renormalized. However,  the squared ANCs are affected by about 0.8$\%$ in both cases. We will keep the same iterative procedure when including the 3N force. 

\begin{figure}[t!]
\centering
\vspace{0.5 cm}
\includegraphics[scale=0.3]{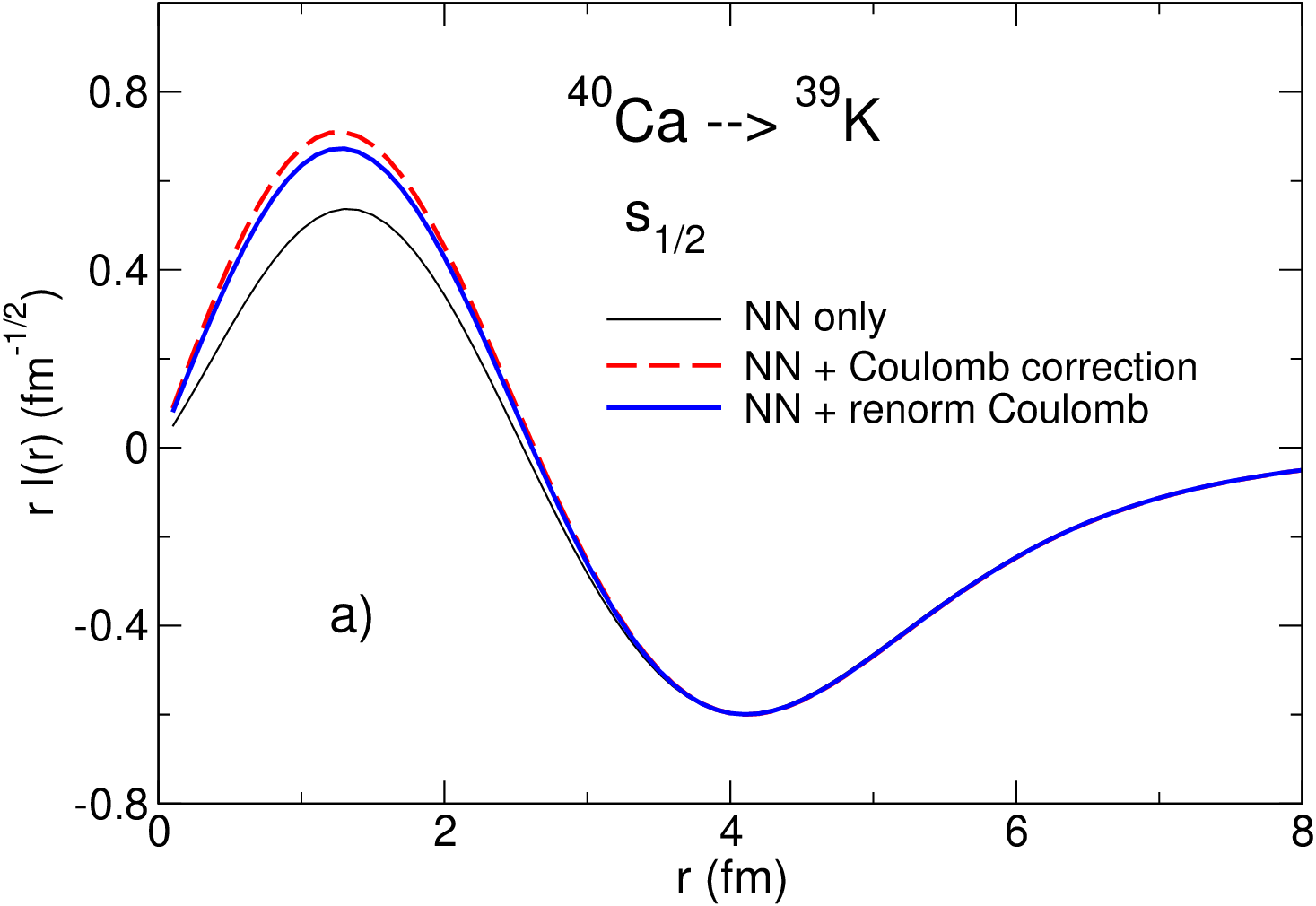}
\includegraphics[scale=0.3]{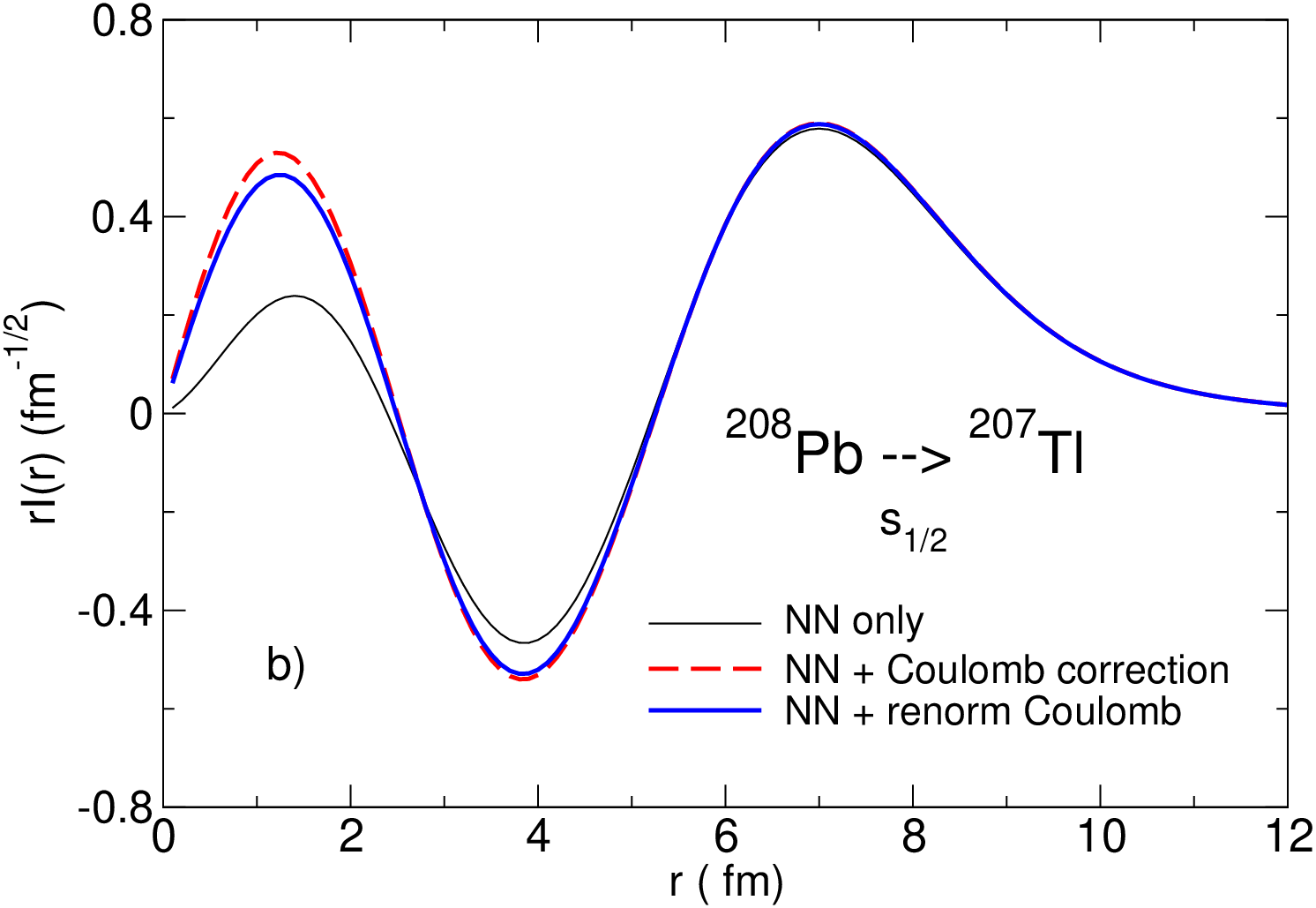}
\caption{The overlap functions $rI_{lj}(r)$ for $s_{1/2}$ proton removal from $^{40}$Ca ($a$) and $^{208}$Pb ($b$) targets calculated in the STA with nuclear NN interaction only (thin line) and with adding the original (dashed line) and renormalised (solid line) Coulomb corrections.
}
\label{fig:CoulST}
\end{figure}

\begin{figure}[h!]
\centering
\vspace{0.5 cm}
\includegraphics[scale=0.3]{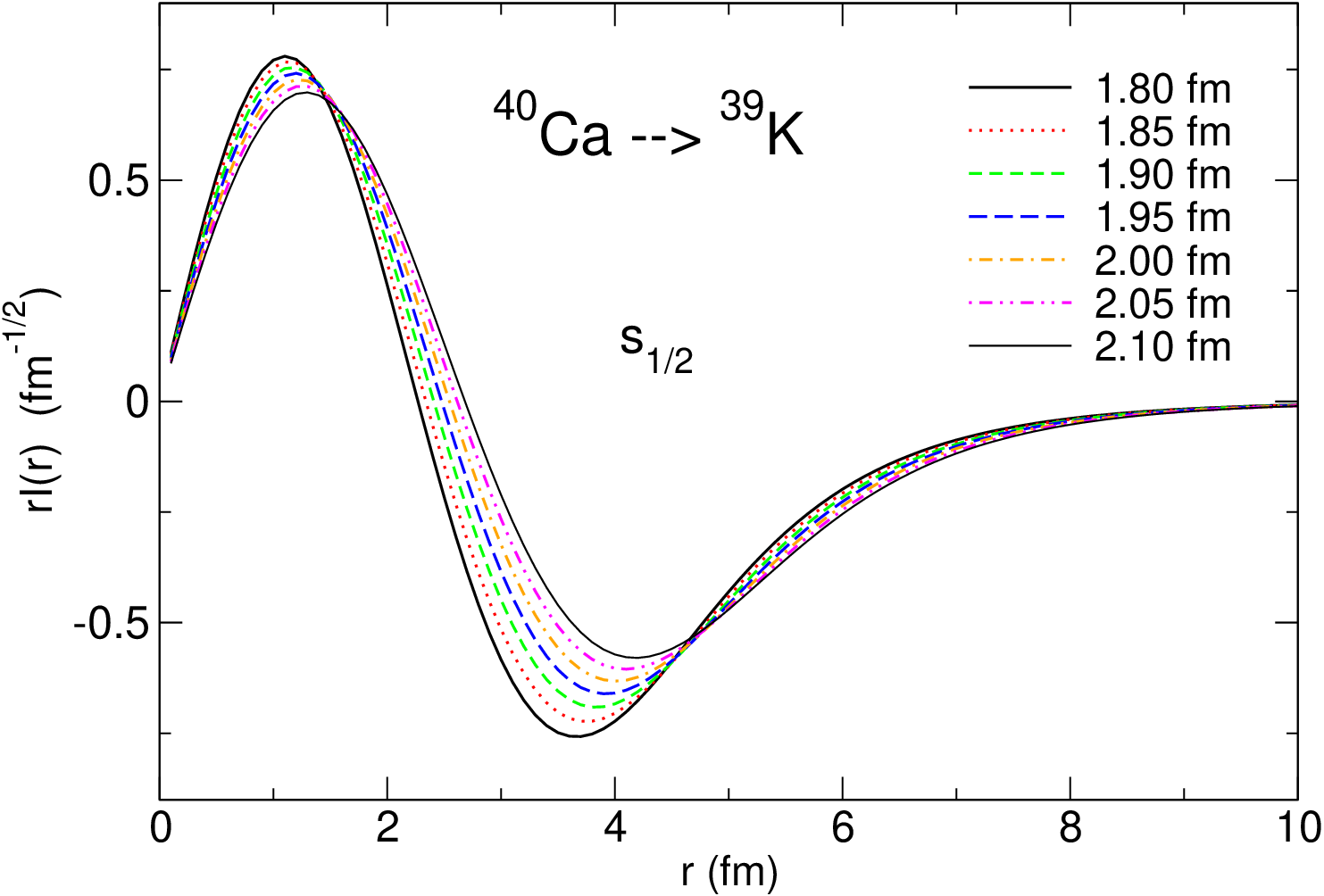}
\includegraphics[scale=0.3]{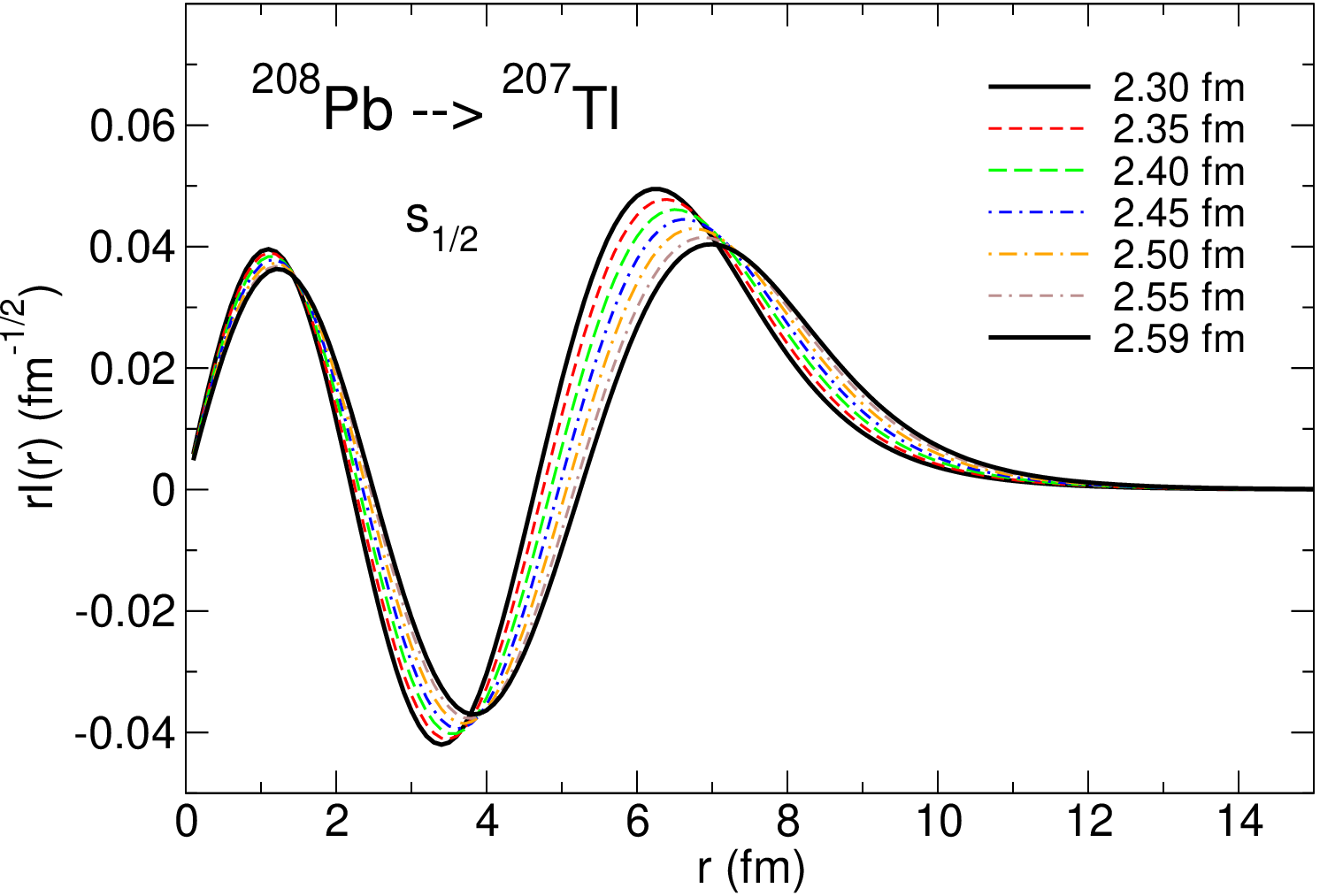}
\caption{The overlap functions $rI_{lj}(r)$ for $s_{1/2}$ proton removal from $^{40}$Ca ($a$) and $^{208}$Pb ($b$) targets calculated in the STA with  NN interaction only for several values of the oscillator radius.
}
\label{fig:oscr}
\end{figure}

Another important input to the STA is the oscillator radius $b$ used to generate harmonic oscillator single-particle wave functions. In previous studies of light $0p$-shell nuclei \cite{Tim09,Tim10,Tim13} this radius has been taken from electron elastic scattering.  Investigation of sensitivity of the overlap functions to the choice of $b$, carried out for $^{16}$O in \cite{Tim10}, has shown that   SFs and ANC can change within a factor of two when  $1.5 <b < 2$ fm. No investigation of $b$-dependence was carried out for $A\geq 16$ nuclei in \cite{Tim11}. The oscillator radius $b$ used there was fixed by the $\hbar \omega = 41A^{-1/3} -25A^{-2/3}$ relation proposed in  \cite{Blo68} which claims that it reasonably reproduces nuclear charge radii. However, analysis of proton scattering on a wide range of  nuclei  
\cite{Dor97,Dor98,Kar01,Kar02,Kar10} suggests that  $b$ can be significantly  smaller. For example, the  $\hbar \omega = 41A^{-1/3} -25A^{-2/3}$ formula gives 2.06 and 2.15 fm for $^{40}$Ca and $^{48}$Ca, respectively, while \cite{Kar10} suggests $b = 1.9 $ fm for both isotopes. A smaller value of $b = 1.869$ fm was used in \cite{Dor97} for $^{48}$Ca and a larger value of $b = 2.0$ fm was used in \cite{Kar01} for $^{40}$Ca.
For $^{208}$Pb, the $\hbar \omega = 41A^{-1/3} -25A^{-2/3}$ prescription gives $ b = 2.59$ fm while \cite{Dor97} and \cite{Dor98} recommend  2.326 fm and $b = A^{1/6} = 2.434$ fm, respectively. Fig. 2 compares overlap functions for $s_{1/2}$ proton removal from $^{40}$Ca and $^{208}$Pb, calculated using NN force only, as a function of oscillator radius. Since both these overlap functions have nodes they  are sensitive to the choice of $b$. This can   affect the spectroscopic factors and, most significantly, the ANCs, which is demonstrated in Fig. 3. While spectroscopic factors change within 30$\%$ and 24$\%$ for $^{40}$Ca and $^{208}$Pb, respectively, the corresponding change in squared ANCs is within 87$\%$ and a factor of 4, which could be completely devastating for peripheral processes.
The sensitivity of the r.m.s. radius of both overlap functions is much smaller.

\begin{figure}[t!]
\centering
\vspace{0.5 cm}
\includegraphics[scale=0.3]{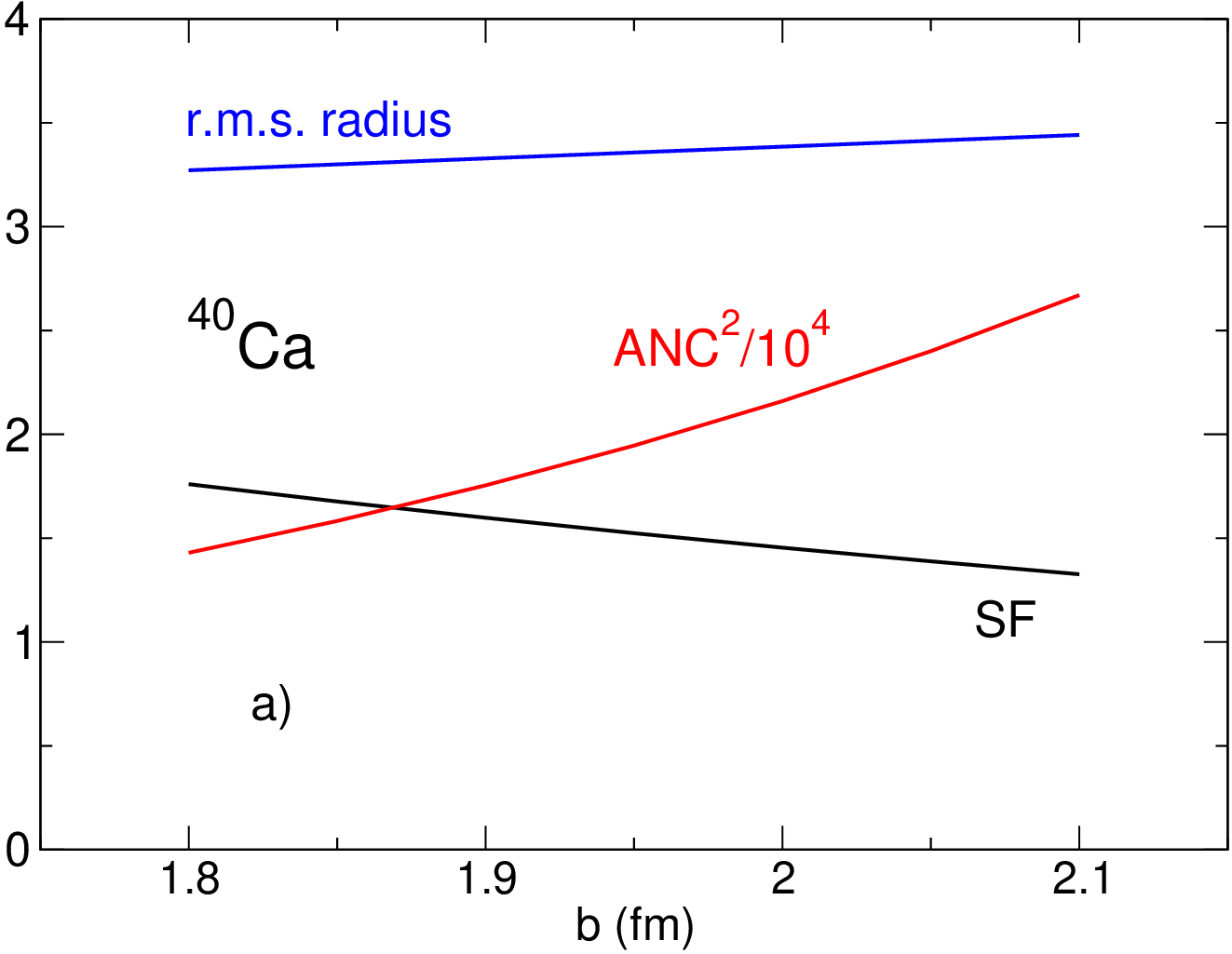}
\hspace{0.8 cm}
\includegraphics[scale=0.3]{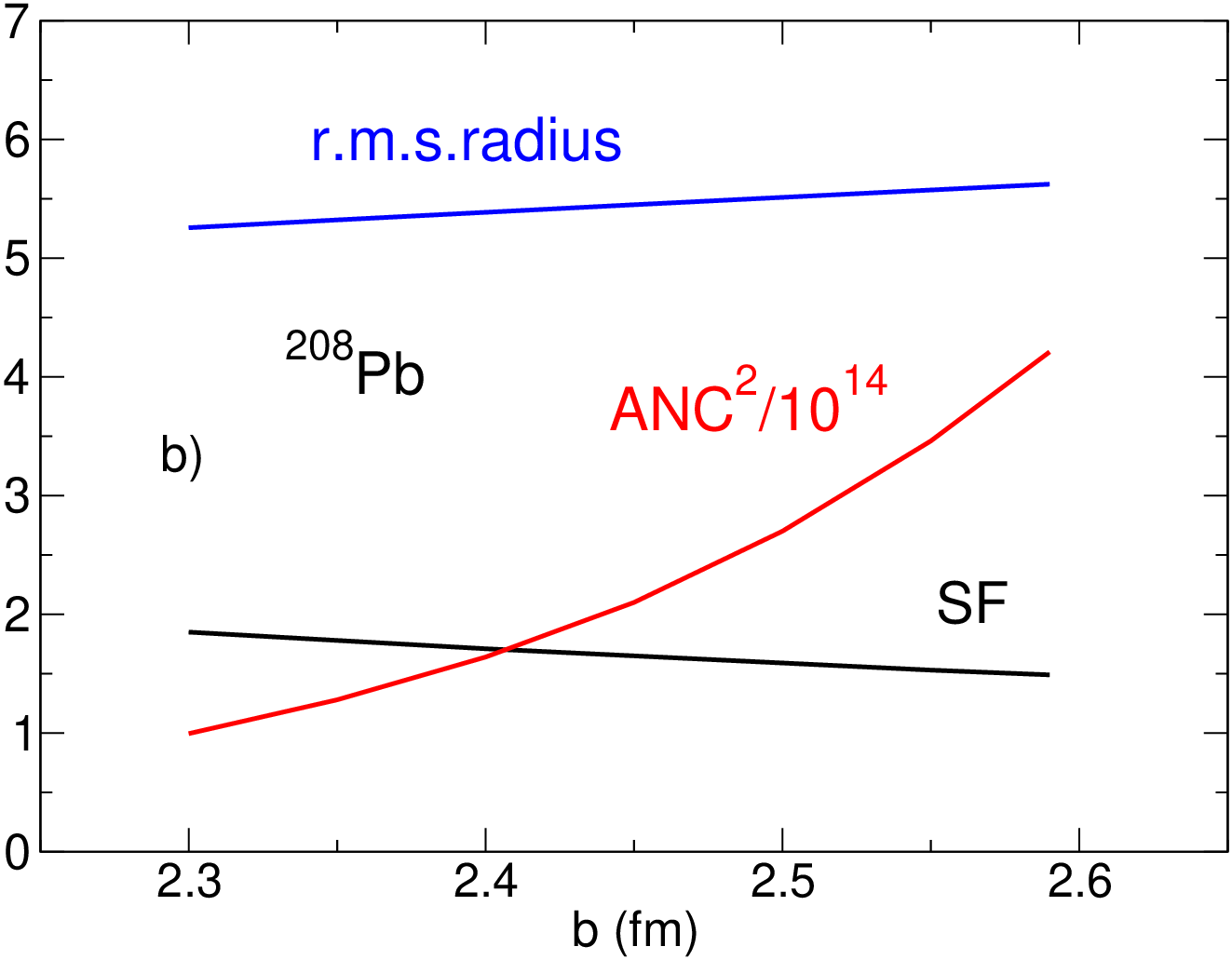}
\caption{The spectroscopic factor (SF), ANC squared (in  fm$^{-1}$) and r.m.s. radius (in fm) for $s_{1/2}$ proton removal from $^{40}$Ca ($a$) and $^{208}$Pb ($b$) targets as a function of the oscillator radius $b$, calculated in the STA with NN interaction only.
}
\label{fig:oscb}
\end{figure}

\section{Numerical calculations with 3N force}

\begin{figure}[t!]
\centering
\vspace{0.5 cm}
\includegraphics[scale=0.3]{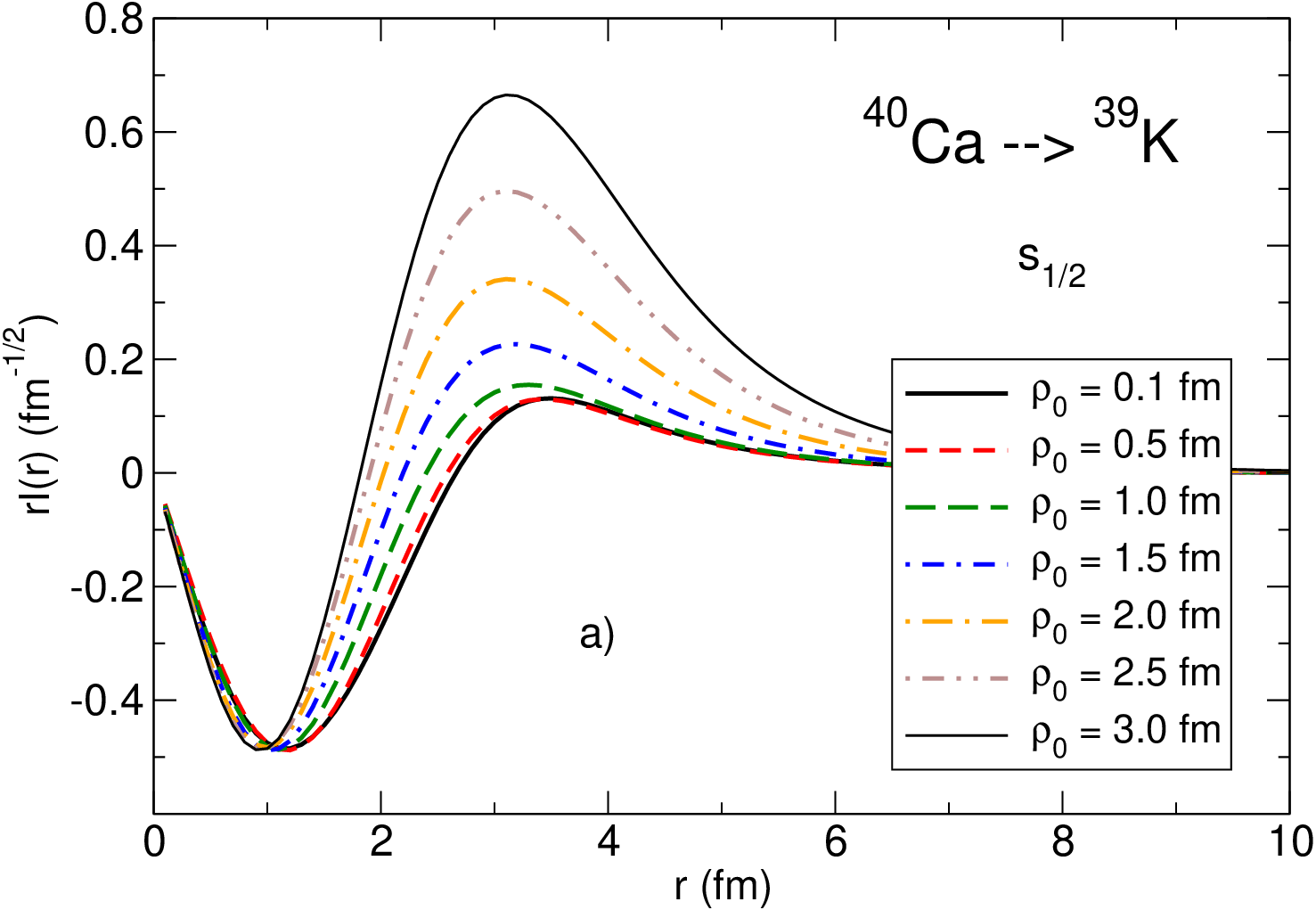}
\includegraphics[scale=0.3]{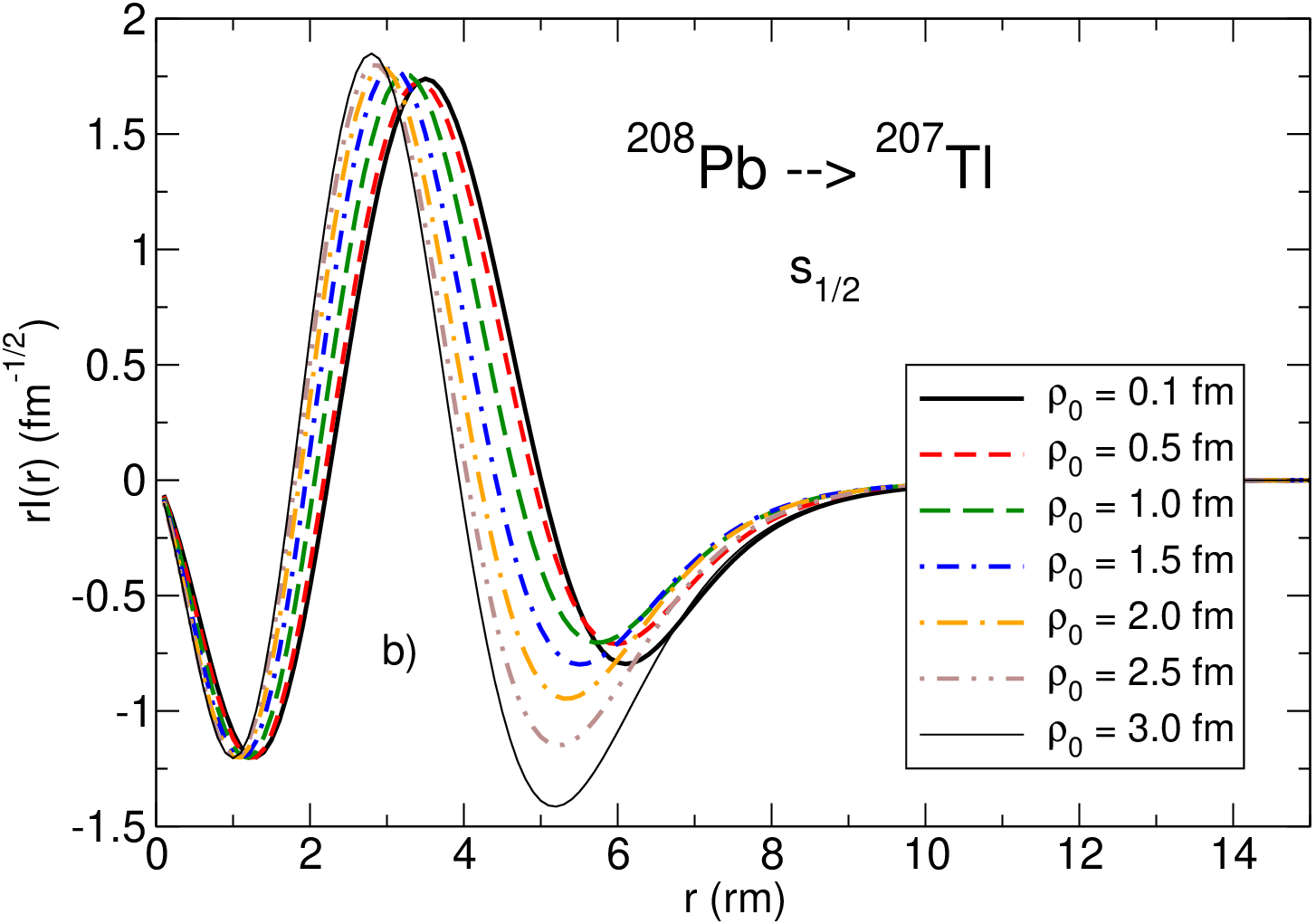}
\caption{The contribution to the overlap function $rI_{lj}(r)$ corresponding to the 3N force $W^{(0)}_{123}$ calculated for $s_{1/2}$ proton removal from $^{40}$Ca ($a$) and $^{208}$Pb ($b$) targets for several ranges $\rho_0$  of the 3N force. The depths $W_0$ are chosen to give the same value for the overlap function in the first maximum for each $\rho_0$.
}
\label{fig:rho0}
\end{figure}

First of all, the dependence of the source term, and the corresponding contribution to the overlap function, on the choice of the range $\rho_0$ of the 3N force was investigated. Figure \ref{fig:rho0} shows  this dependence for the case of the $s_{1/2}$ proton removal from $^{40}$Ca and $^{208}$Pb, demonstrated for the contribution from the $W^{(0)}_{123}$ term only. The overlap functions calculated with different $\rho_0$ were normalized to have the same value in the first maximum. For both nuclei increasing $\rho_0$ leads to a larger second maximum and moves the position of the first node towards smaller $r$. For $^{208}$Pb, the third maximum first decreases at $0 < \rho_0 < 0.5$ fm but then increases  for $\rho_0> 0.5$ fm. For both nuclei the nodes of $I_{lj}(r)$ are closer to the origin than those  of the overlaps  generated by the NN force only. This means that if the NN force does not reproduce the magnitude of the asymptotic part of $I_{lj}(r)$, corresponding to  experimentally determined ANCs, the empirical 3N force needed to correct this magnitude would have a large range $\rho_0$.

\begin{figure}[t!]
\centering
\vspace{0.5 cm}
\includegraphics[scale=0.3]{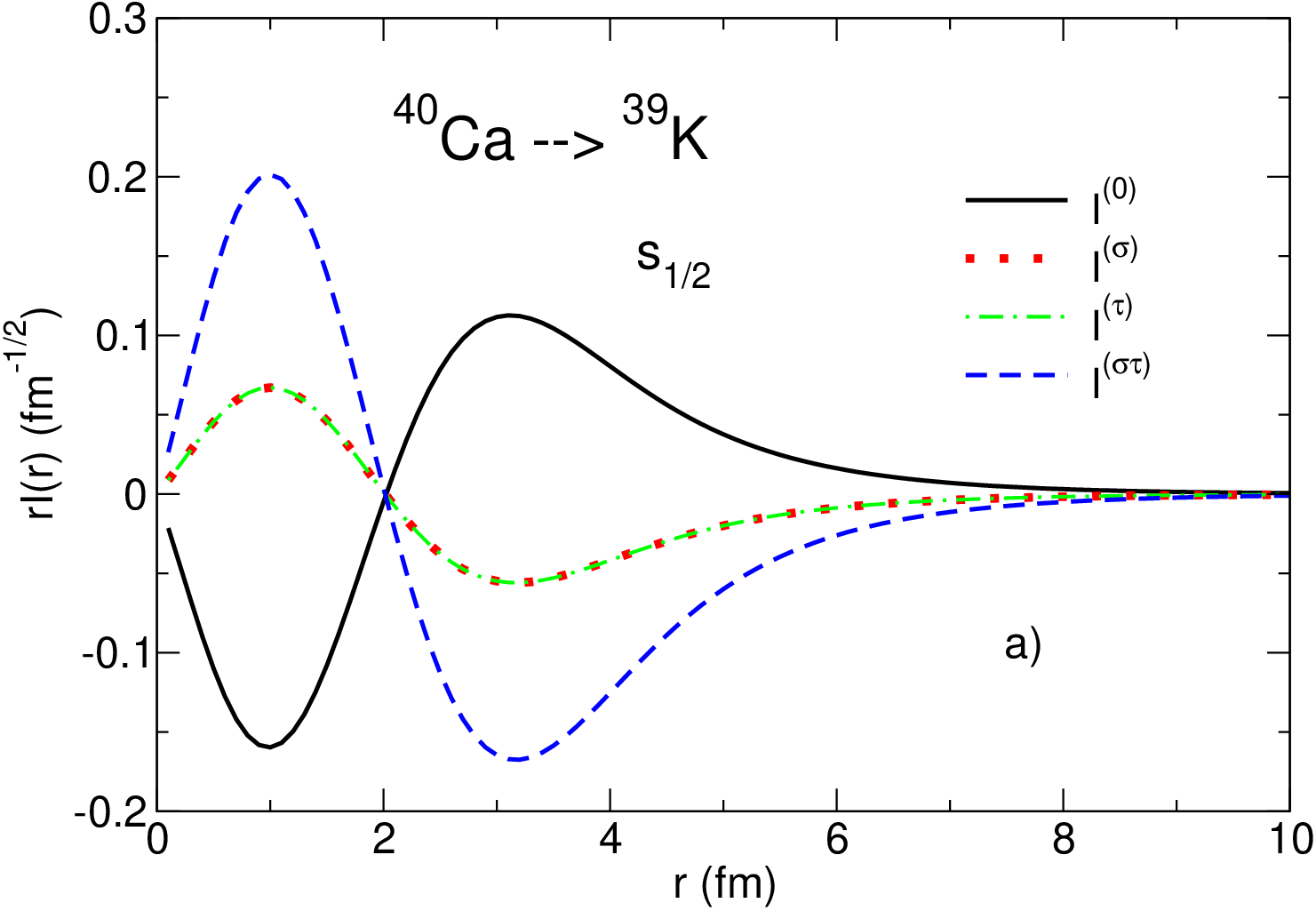}
\includegraphics[scale=0.3]{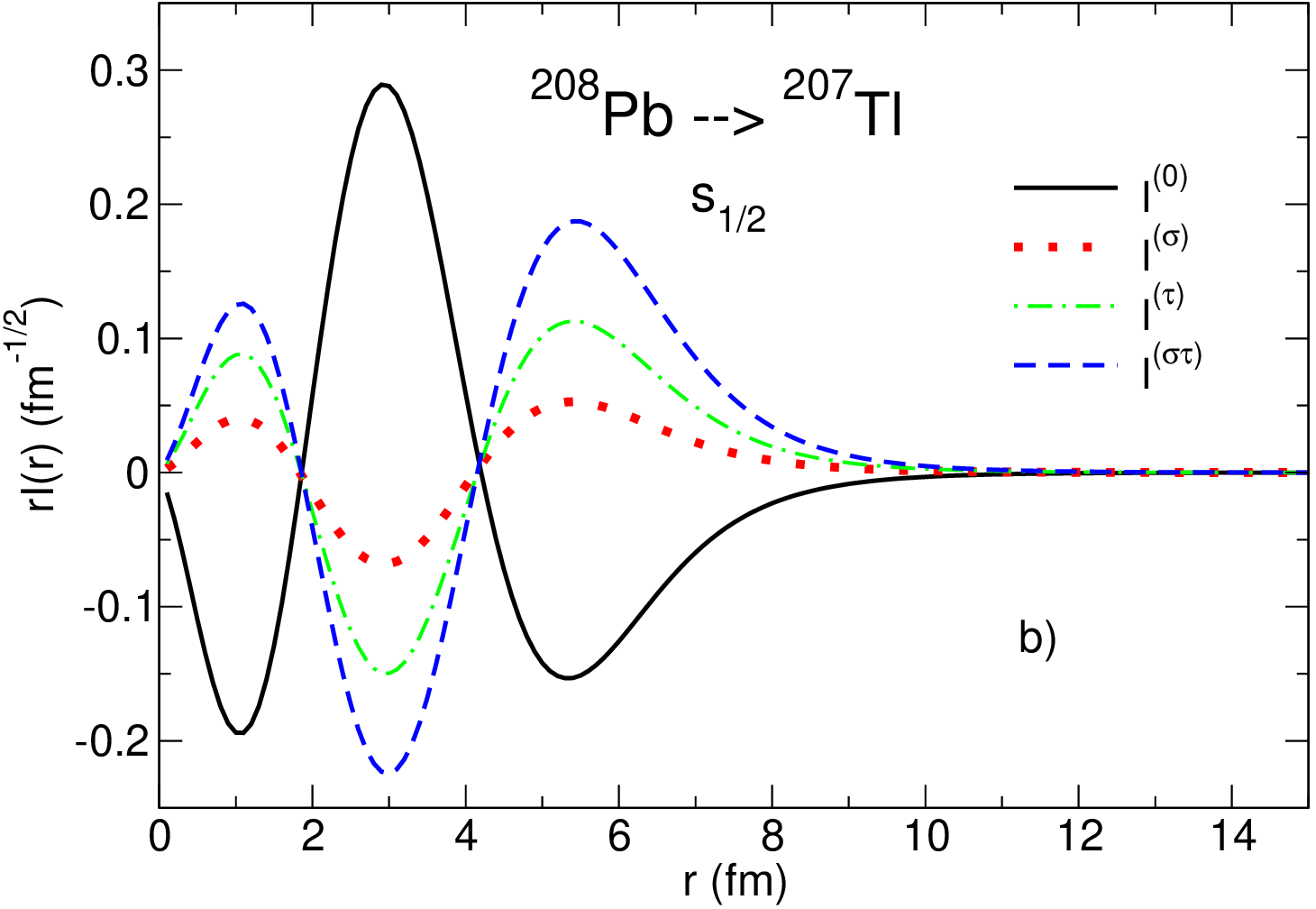}
\caption{The contribution to the overlap function corresponding to the 3N force components $W^{(0)}$, $W^{(\sigma)}$, $W^{(\tau})$ and $W^{(\sigma\tau)}$, calculated for $s_{1/2}$ proton removal from the $^{40}$Ca ($a$) and $^{208}$Pb ($b$) targets using $\rho_0$ = 2 fm.  The oscillator radii employed are $ b = $1.90 fm and 2.38 fm for $^{40}$Ca and $^{208}$Pb, respectively.
}
\label{fig:I0st}
\end{figure}

Next, the relative contributions from different components of the 3N force have been studied. Figure \ref{fig:I0st} shows these contributions calculated for $s_{1/2}$ proton removal from $^{40}$Ca and $^{208}$Pb using $\rho_0=2$ fm   and the depth of  1 MeV for all four components of the 3N force. For $^{40}$Ca the contributions from $W^{(\sigma)}$ and $W^{(\tau)}$ are the same, being exactly one-third of the contribution from $W^{(\sigma\tau)}$. The same has been observed for removals   of nucleons with other $l$ and $j$ involving $^{16}$O and $^{40}$Ca. These nuclei have a well defined value of the spin $S=0$  (associated with the operator $\vec{S} = \sum\limits_i \vec{s}_i$ and not to be confused with the total angular momentum $J$) and isospin $T=0$. For other double-closed shell nuclei considered here, where $S$ is not a good quantum number, such relations between  the corresponding components do not hold any longer. For example, for $^{208}$Pb the proportions between the values of the first, second and third maximum are different for all four components of the 3N force, which could be helpful when  fitting it to get the correct positions of the nodes of the overlap.  In all cases considered the contribution from $W^{(0)}$ has a different sign to those coming from $W^{(\sigma)}$, $W^{(\tau)}$ and $W^{(\sigma\tau)}$ and is not proportional to them.

%well-defined value of the  operator $S^2 = \left(\sum_i \vec{s}_i\right)^2$.and isospin

 %For nucleon addition and removal from $^{16}$O and $^{40}$Ca, where the total spin and isospin are well defined quantities equal to zero, we obtained $ 3S_{\sigma}^{(3N)}(r) = 3S_{\tau}^{(3N)}(r) = S_{\sigma\tau}^{(3N)}(r)$. The $S_0^{(3N)}(r)$. The corresponding 3N contributions $I_0^{(3N)}(r)$,  $I_{\sigma}^{(3N)}(r)$,  $I_{\tau}^{(3N)}(r)$ and $I_{\sigma \tau}^{(3N)}(r)$ to the overlap integral  enter in the same proportions. All other heavier double-closed shell nuclei $^{48}$Ca, $^{56}$Ni, $^{208}$Pb do not have a well-defined value of the  operator $S^2 = \left(\sum_i \vec{s}_i\right)^2$. Adding or removing a nucleon to them results in an $S_{\sigma}^{(3N)}(r)$ term that differs from $S_{\tau}^{(3N)}(r)$ by some factor, being still proportional to it,  but $S_0^{(3N)}(r) = S_{\sigma\tau}^{(3N)}(r) =  3S_{\tau}^{(3N)}(r)$ is still valid. This means that for any individual nucleus, considered in this work, only one component of the 3N force is sufficient to be included into consideration.

%Figure xxx shows $rI^{3N}(r)$ calculated for $s_{1/2}$ proton removal from $^{40}$Ca and $^{208}$Pb with several valued of range $\rho_0$ of the 3N force. One can see that for $^{40}$Ca the first maximum is about xxx times stronger than the second maximum for all ranges $\rho_0$. In contrast, the corresponding overlap, calculated with 2N force only has different proportions between the two maxima (see Fig. zzz). This means that adding 3N effective force will mainly affect  the overlap at small $r$ and, therefore, will be more important for calculations of the spectroscopic factors rather than ANCs. The same is true for the $^{208}$Pb, where... This is true for all other cases considered here as well. Therefore, if 3N is meant to fitted to overlap integrals determined empirically from experimental data, it is important to use the oscillator radius that gives the best possible ANC when used in calculations with 2N force only.

\begin{figure}[t!]
\centering 
\vspace{0.3 cm}
\includegraphics[scale=0.202]{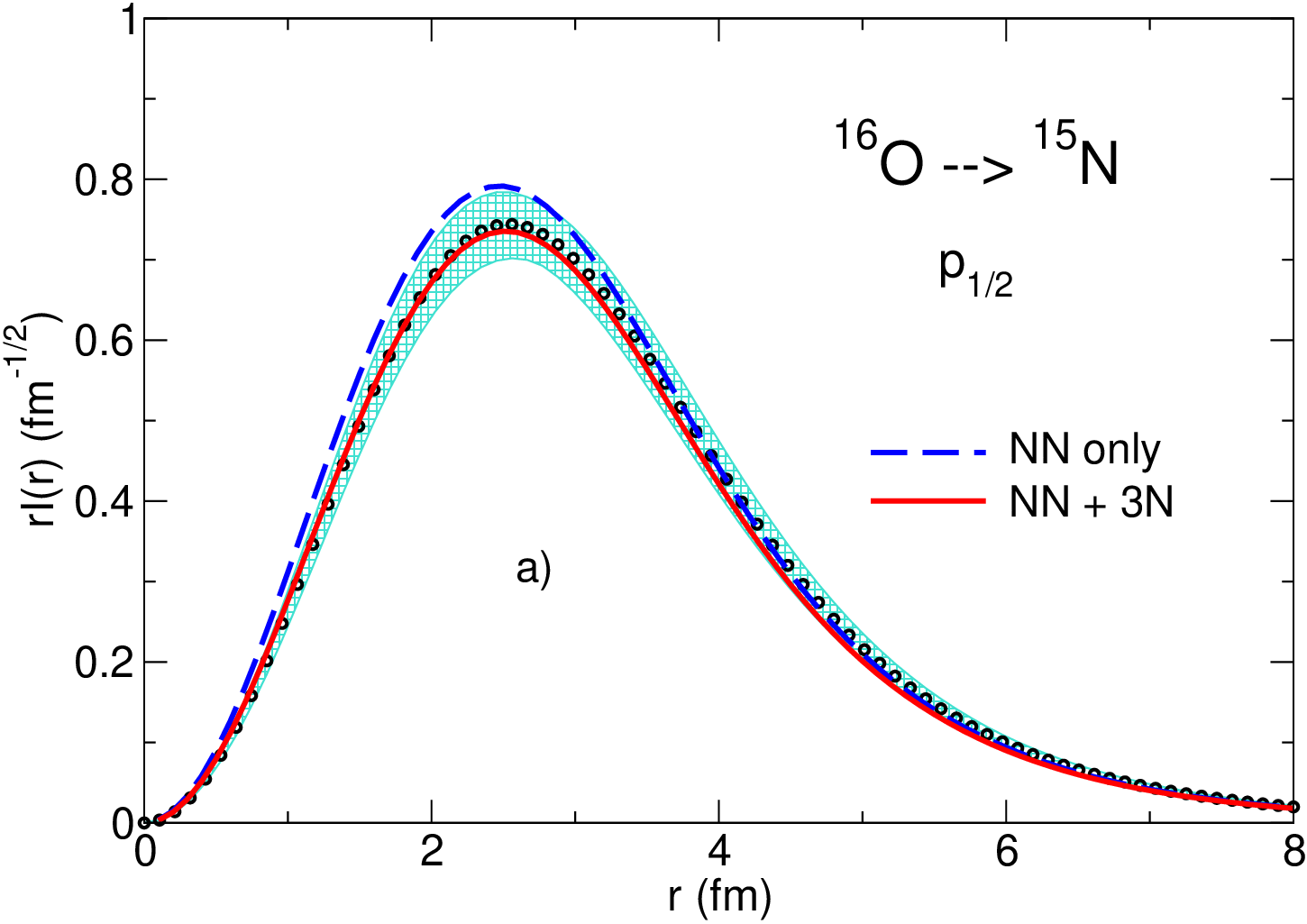}
\includegraphics[scale=0.202]{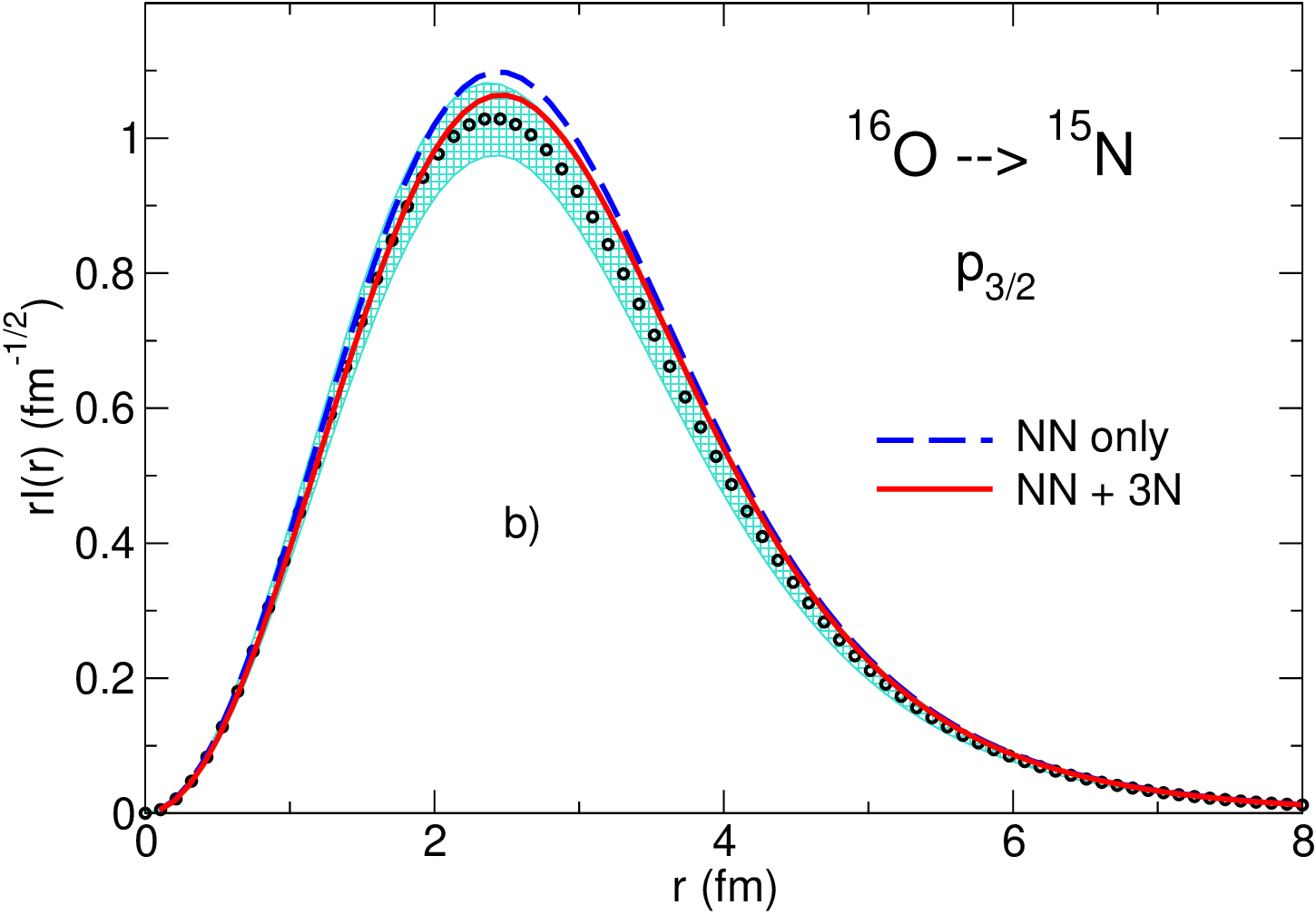}
\vspace{0.8 cm}
\includegraphics[scale=0.202]{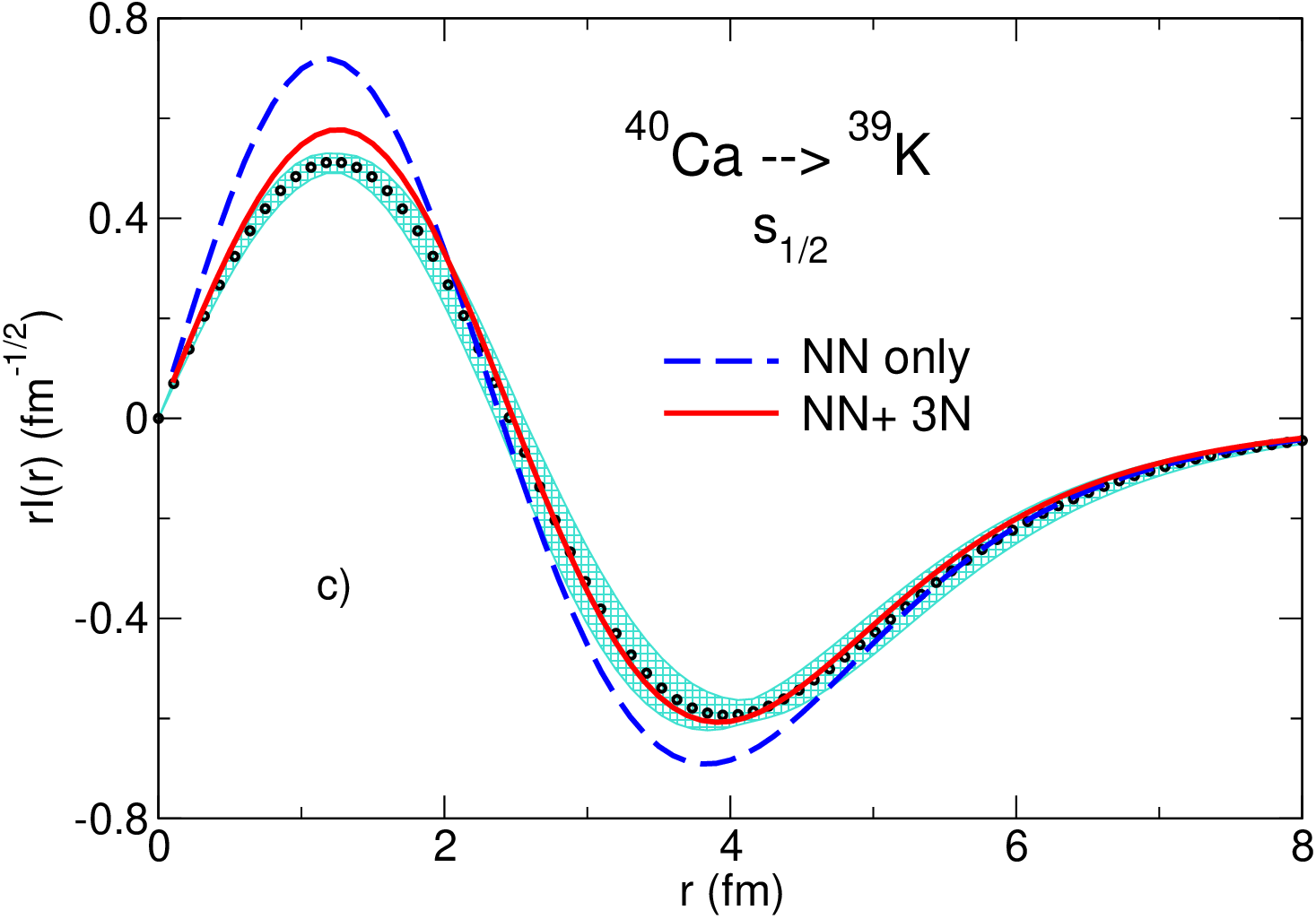}
\includegraphics[scale=0.202]{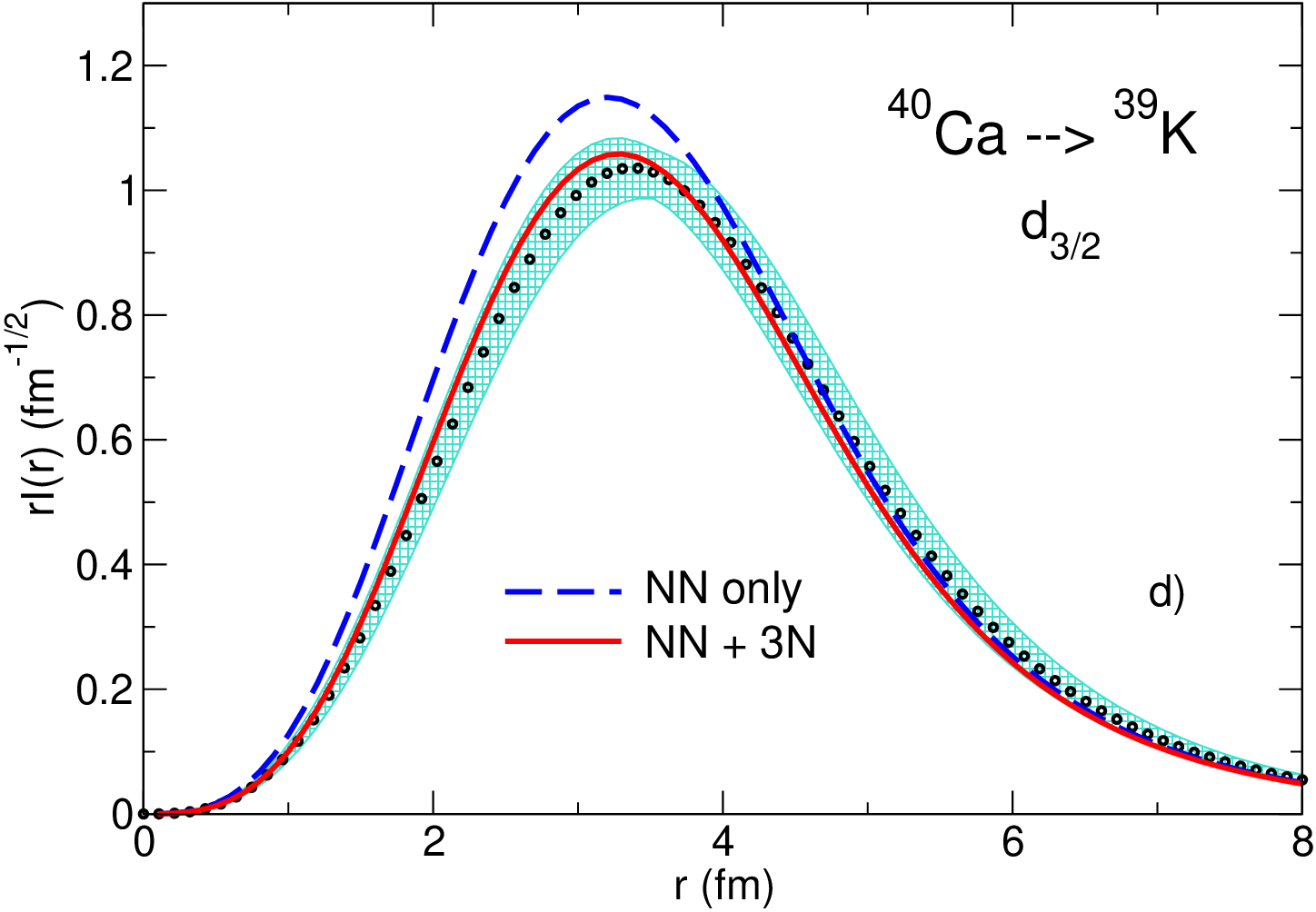}
\vspace{0.6 cm}
\includegraphics[scale=0.202]{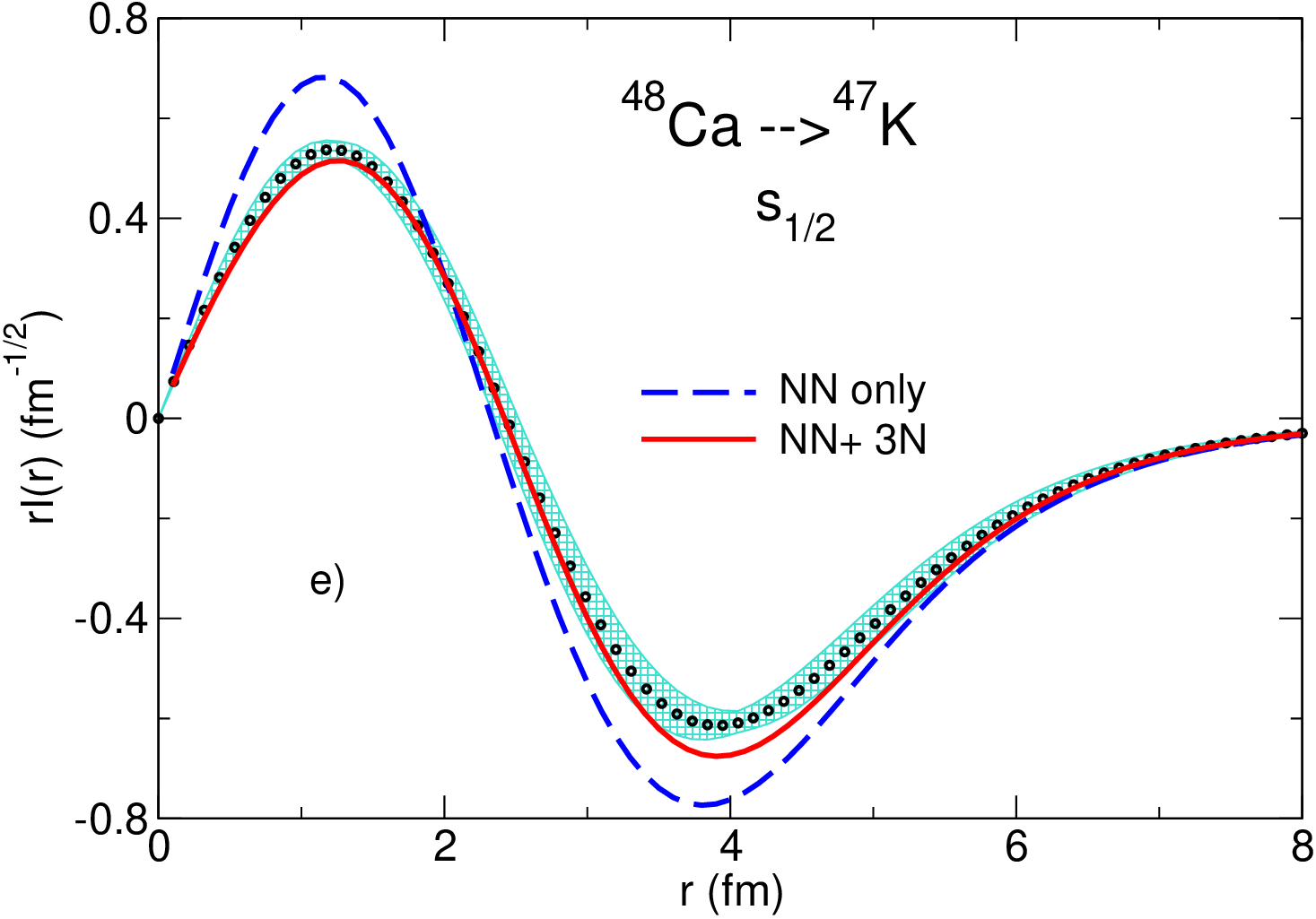}
\includegraphics[scale=0.202]{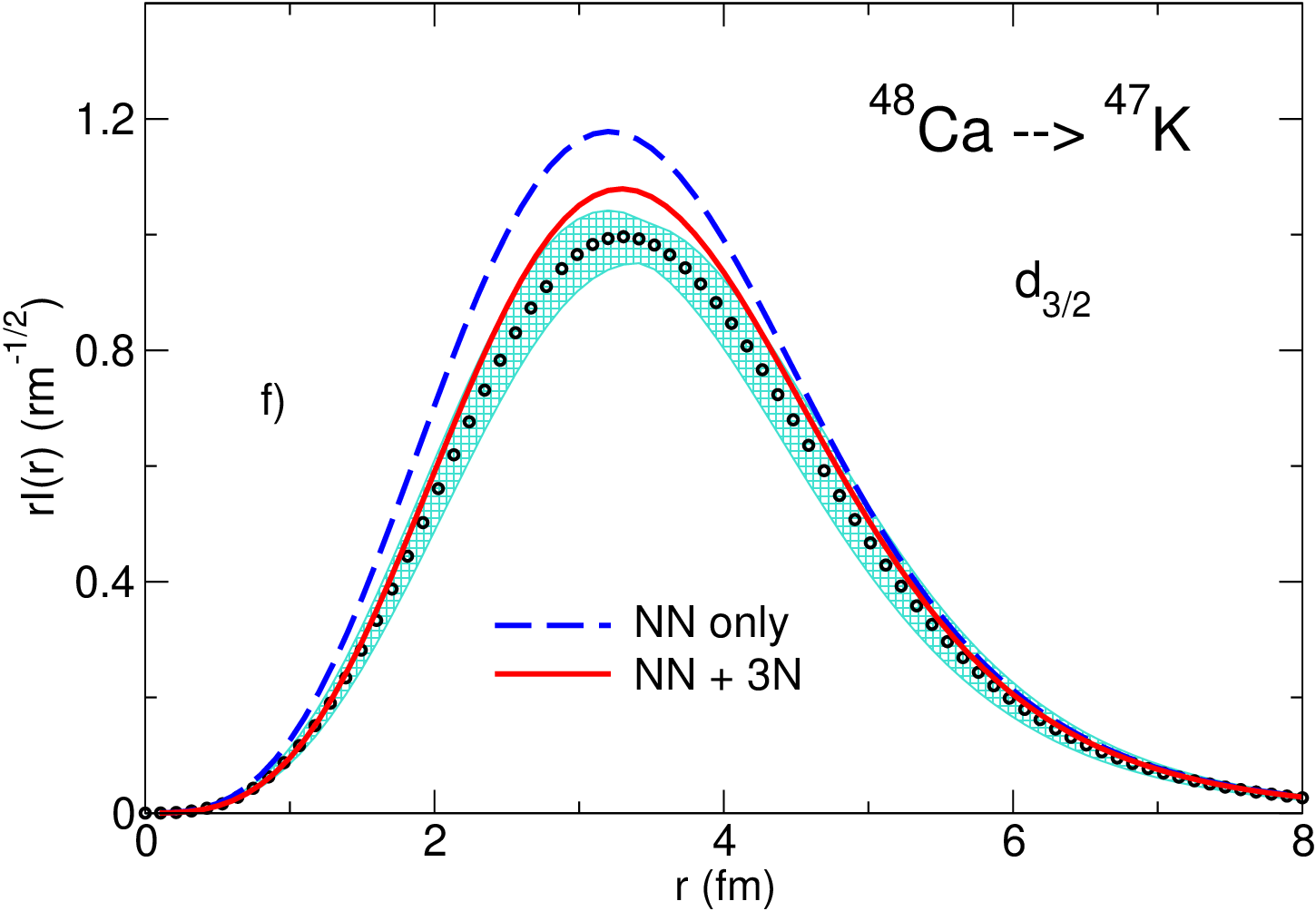}
\vspace{0.6 cm}
\includegraphics[scale=0.202]{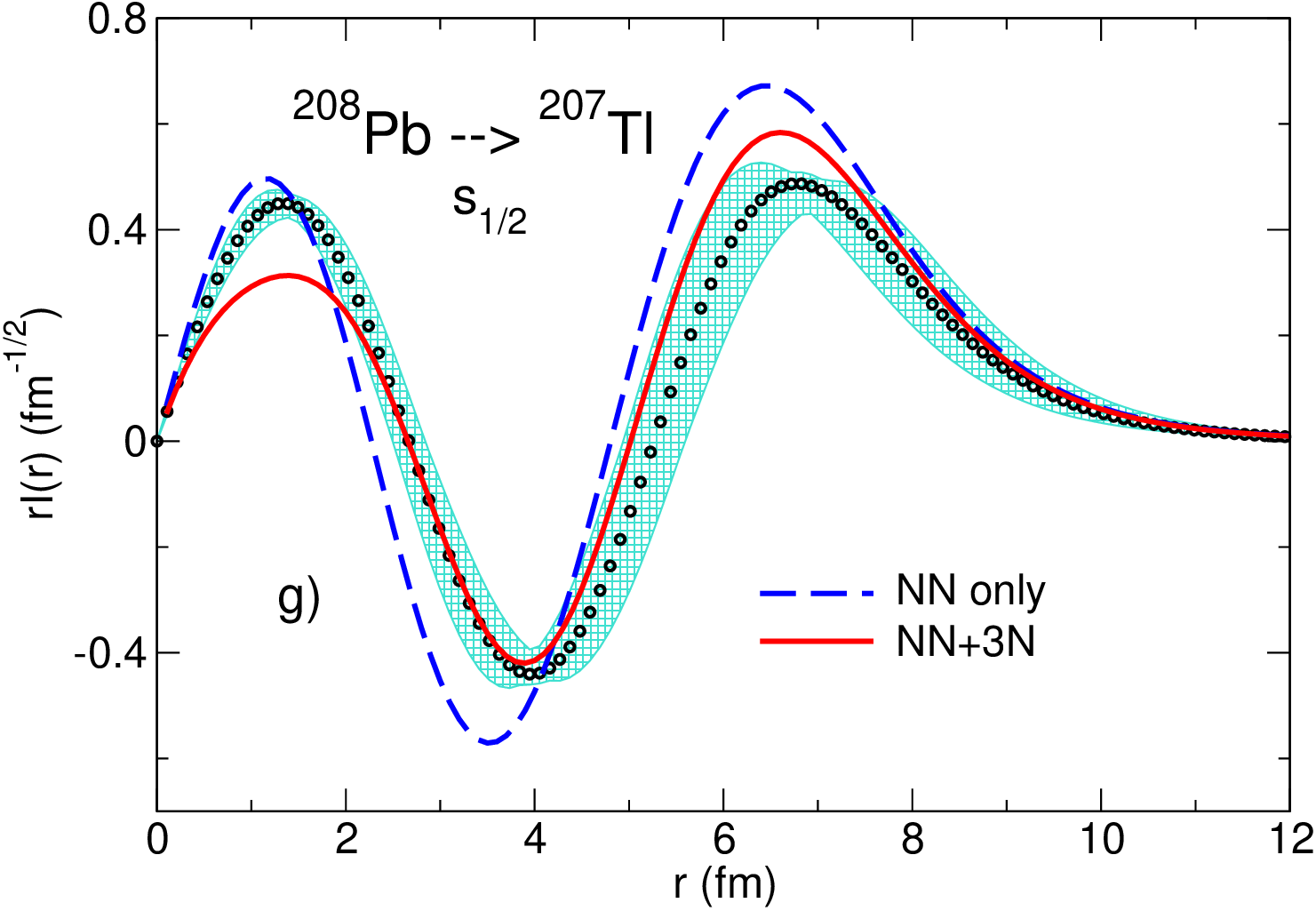}
\includegraphics[scale=0.202]{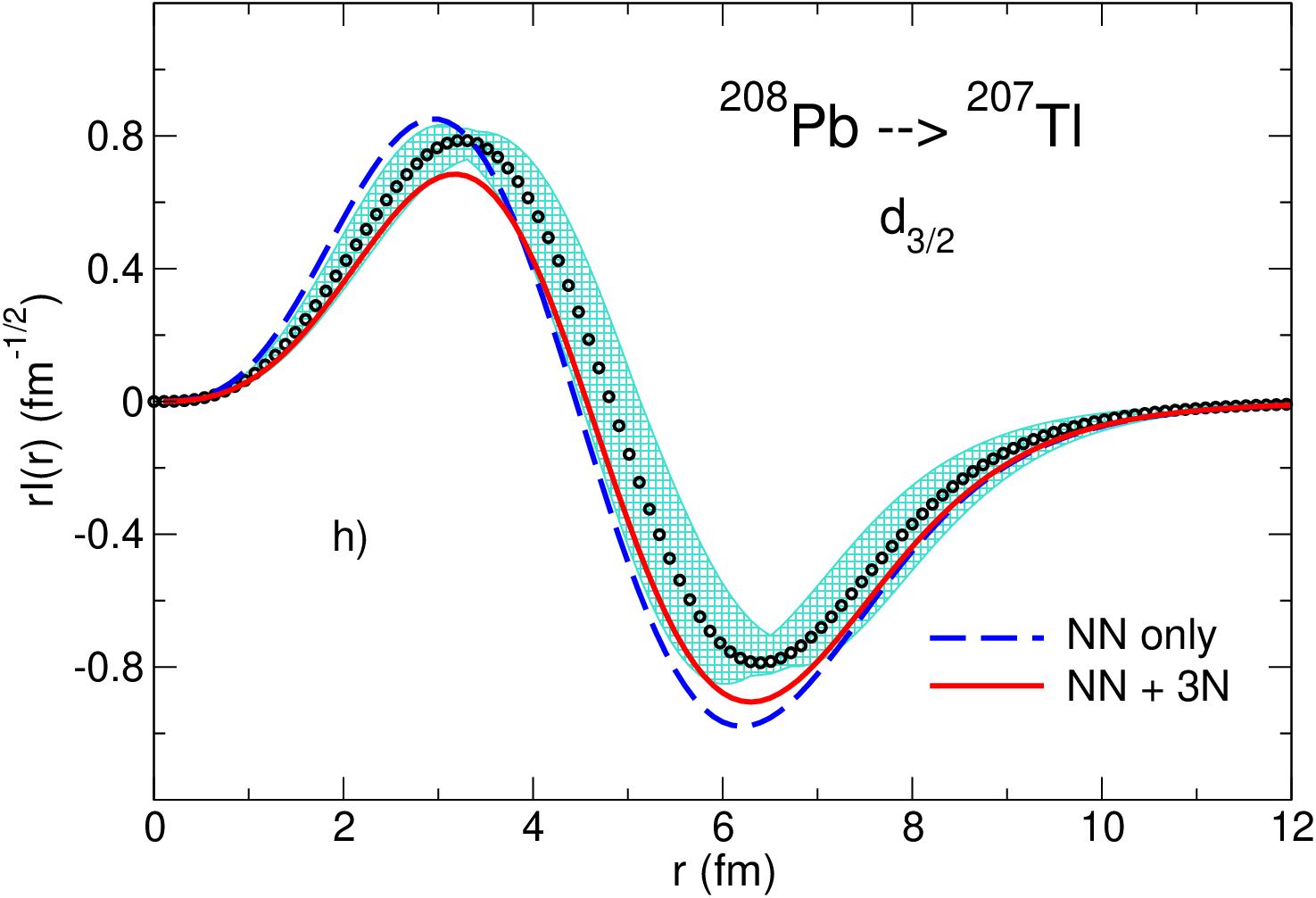}
\vspace{0.6 cm}
\includegraphics[scale=0.202]{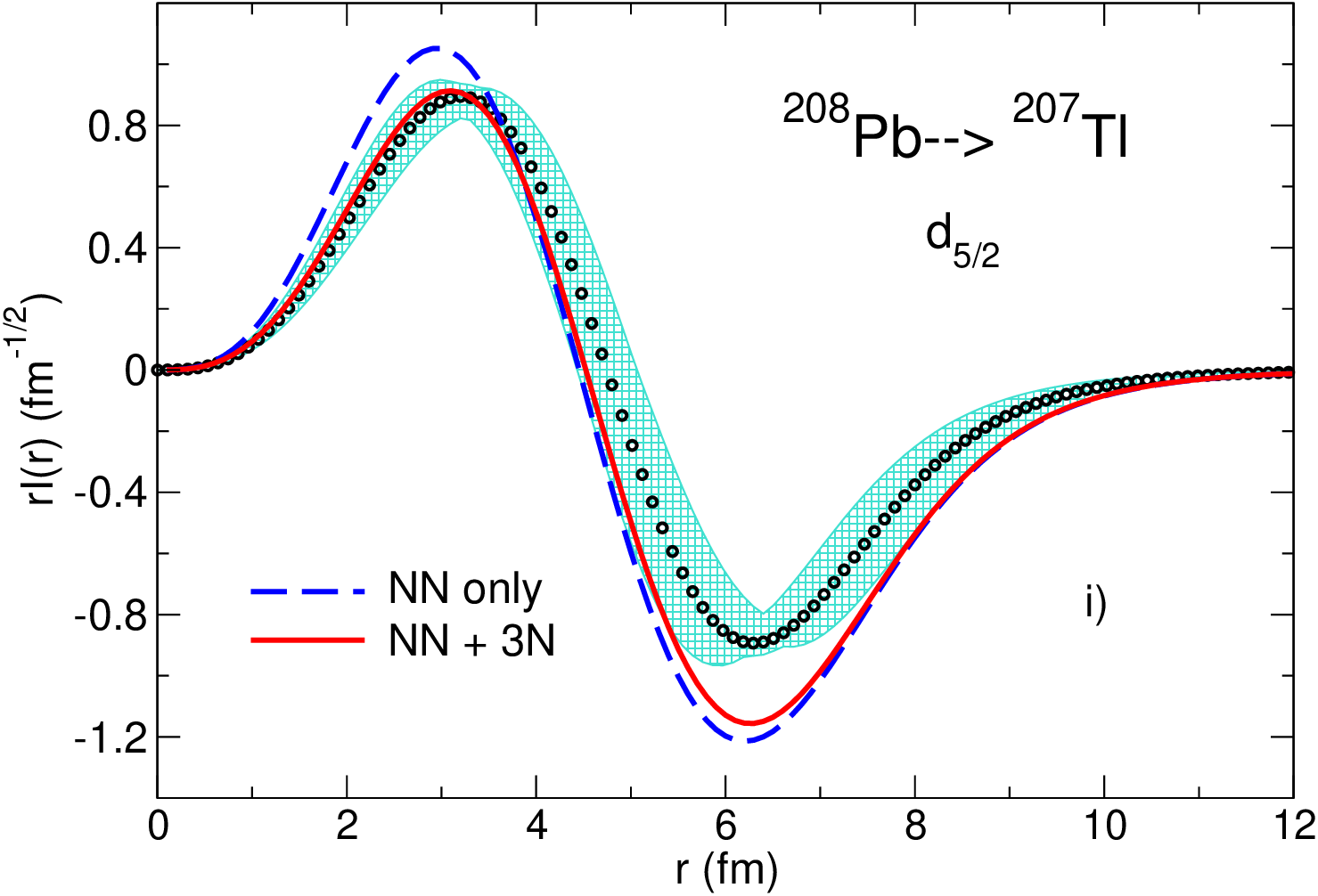}
\includegraphics[scale=0.202]{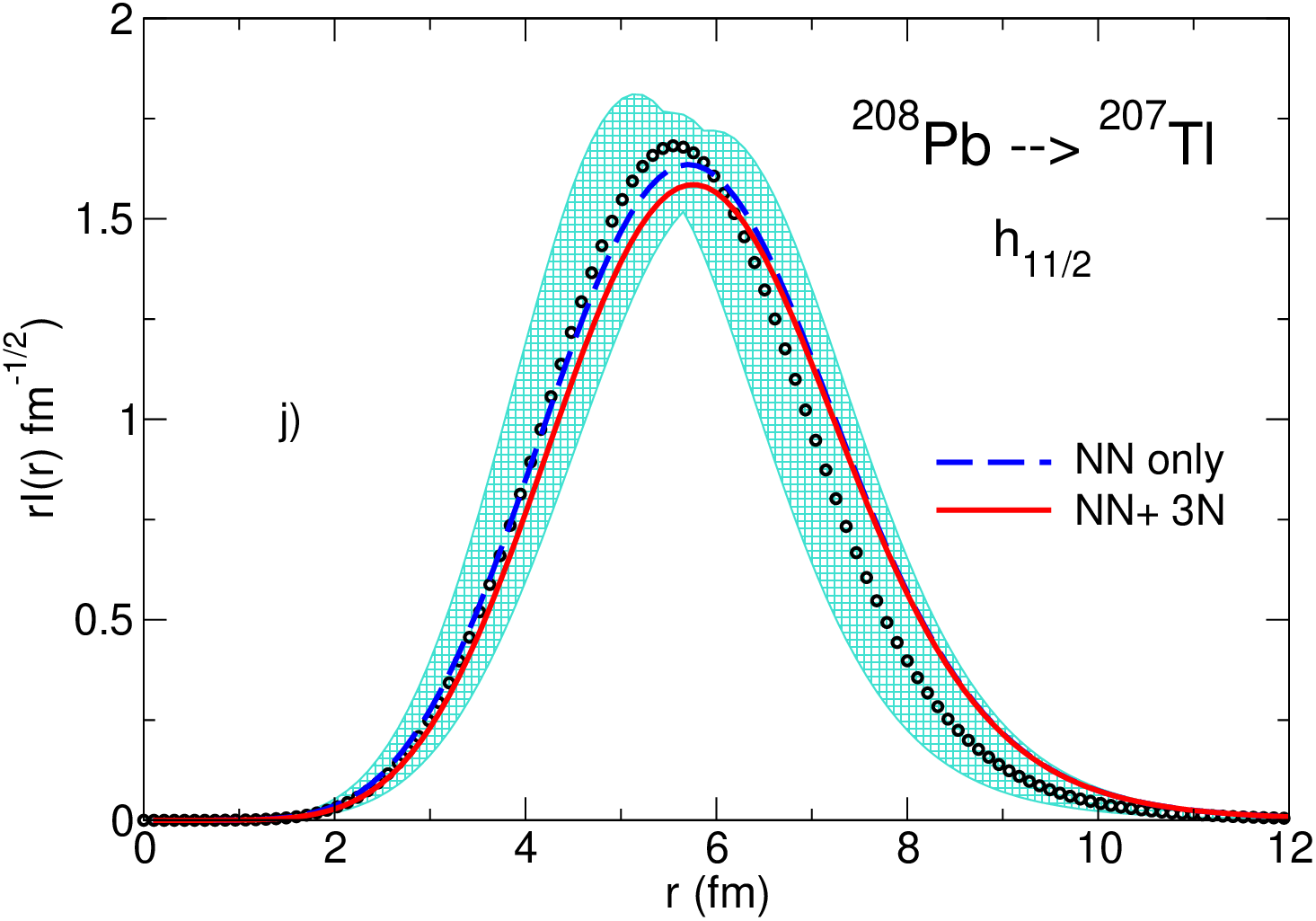}
\caption{One-proton removal  overlap functions for  $^{40}$Ca ($a,b)$, $^{48}$Ca ($c,d$) and $^{208}$Pb $(e,f,g,h)$, calculated with NN force only and with including the 3N force having   $W_0= 1$ MeV and $\rho_0 = 2$ fm.
}
\label{fig:2N3N}
\end{figure}

As a next step  it was investigated how strong the 3N force should be to give reasonable predictions for the overlap functions for proton knockout from double-magic nuclei $^{16}$O, $^{40,48}$Ca and $^{208}$Pb. Phenomenological overlaps determined from the $(e,e'p)$ reactions on these targets  are available in \cite{Kra01} 
%, which is, perhaps,  the only systematic work devoted to consistent study of a wide range of the overlaps within the same reaction theory.
%It should be noted though that    recently  one-nucleon removal functions  were extracted from analysis of  elastic nucleon scattering from $^{40,48}$Ca and $^{208}$Pb   within dispersive optical model [refs]. They confirm the spectroscopic factors from \cite{Kra01}, however,   the single-particle energies of these functions differ from experimental nucleon separation energies, affecting their asymptotic decrease, which makes such overlaps unusable in one-nucleon transfer calculations.
where they   are represented by the single-particle wave functions, satisfying two-body Schr\"odinger equation with experimental separation energies, times the spectroscopic amplitudes. Such a representation guarantees that the  overlaps have a correct asymptotic decrease. The radius $r_0$ and diffuseness $a$ of the potential well used to calculate the single-particle wave functions are given in Tables 1 and 3 of \cite{Kra01}, together with the spectroscopic factors $S_{lj}$. %The depth of the two-body potential well has been fitted to proton separation energies. Although this procedure gives the  overlaps with a correct asymptotic decrease, 
Both $r_0$ and $S_{lj}$ are given together with their uncertainty ranges. The corresponding   overlaps  also have uncertainties that can be large, % in its magnitude   could be very large,
as  seen in figures \ref{fig:2N3N}a to \ref{fig:2N3N}h. In these figures the overlaps corresponding to the mean values of the parameters $r_0$ and $S_{lj}$ are shown by open circles while all other possible values are collected in a band. 
%because in \cite{Kra01}  the radii $r_0$ and spectroscopic factors are given with experimental uncertainties, which affect $I^{\exp}_{lj}$, especially in the asymptotic region. That region was not important for $(e,e'p)$ reactions but is crucial for transfers. It is not clearly explained in \cite{Kra01} how uncertainties in $r_0$ and spectroscopic factors are related.  
%Because of the large uncertainties of phenomenological overlaps no attempts to fit  the 3N force were made. Instead, adding contributions calculated with a fixed value of $\rho_0$ and a slightly adjusted  depth $W_0$ are shown to give an idea about general influence of this force.

In most cases shown in Fig. \ref{fig:2N3N} the overlaps calculated with NN force only  are outside the phenomenologically determined bands, which is partially due to the choice of the oscillator radius.  The calculations with NN force used smaller oscillator radii than those employed in all previous STA publications  and they were based on the proton scattering work of \cite{Dor97,Dor98,Kar01,Kar02,Kar10} being 1.8 fm for $^{16}$O, 1.90 fm for $^{40,48}$Ca and 2.38 fm $^{208}$Pb. Larger oscillator radii   lead to larger  overlaps   outside the nuclear interior so that unrealistically large ranges of the 3N force are needed to bring them down. It was found out that adding only one component of the 3N force, $W^{(0)}$,  with a depth  $W_0 = 1$ MeV and a range $\rho_0 = 2$ fm is sufficient to get an improved description of the phenomenological  overlaps for all considered final states populated by proton knockout from $^{16}$O, $^{40,48}$Ca and $^{208}$Pb, which is shown in Fig.  \ref{fig:2N3N}.  An alternative single-term 3N potential with $W^{\sigma\tau}_0 = -0.9$ MeV and $\rho_0 = 2$ fm gives  a similar quality (or slightly better) description of the phenomenological overlaps for $^{40,48}$Ca and $^{208}$Pb but a worse description for $^{16}$O. It should be noticed that for any individual overlap function it is possible to tune the 3N force to locate $I_{lj}(r)$  within the limits of the phenomenological band. However, there is no point is making such an effort at present  since the method itself would lose predictability. More important is to understand if a universal effective 3N force exists that would be applicable to all nuclei. However, this is a task for the future.

\begin{table}[h]
\caption{ The STA
spectroscopic factors for the overlap functions $\la A | A-1\ra $ obtained  with NN force only ($S_{lj}^{NN}$) and  with including 3N force ( $S_{lj}^{(NN+3N)}$). The spectroscopic factors obtained from analysis of $(e,e'p)$ reaction for $^{16}$O, $^{40,48}$Ca and $^{208}$Pb are shown in the last column.
 Also shown are the quantum numbers $lj$ of the removed nucleon,  
 the excitation energy $E_x$ 
 of $A-1$, the separation energy $\varepsilon$ of the removed nucleon (both in MeV) and
the spectroscopic factor $S^{\rm SM}_{lj}$ of the $0\hbar \omega$ shell model. For $^{16,24}$O, $^{40,48}$Ca, $^{56}$Ni and proton removal from $^{132}$Sn the shell model spectroscopic factors include correction for centre-of-mass motion.
 } 
\begin {center}
\begin{tabular}{ p{1.0cm} p{1.0 cm} p{1.1 cm} p{1.1 cm} p{1.5 cm} p{1.5 cm}
p{1.5 cm} p{1.7 cm} p{1.5 cm}}
%\begin{tabular}{ ccccccccccc}
\hline
\hline
 $A$ &  $A$$-$1 & $lj$ & $E_x$ & $\varepsilon$ & 
   $S^{\rm SM}_{lj}$ & $S_{lj}^{(NN)}$  &  $S_{lj} ^{(NN+3N)}$ & $S_{lj}^{(e,e'p)}$   
 \\
 \hline
$^{16}$O & $^{15}$N & $p_{1/2}$ & 0.0 & 12.13 &   2.133 & 1.42 & 1.23 & 1.27(13) \\
         &          & $p_{3/2}$ & 6.32 & 18.45 &   4.267 &2.56 & 2.40 & 2.25(22) \\
$^{16}$O & $^{15}$O & $p_{1/2}$ & 0.0  & 15.66 &   2.133 & 1.38 & 1.17 \\
         &          & $p_{3/2}$ & 6.18 & 21.84 &   4.267 & 2.53 &2.35 \\
$^{24}$O & $^{23}$N & $p_{1/2}$ & 0.0 & 26.6 &   2.087 & 1.29 & 1.04  \\
$^{24}$O & $^{23}$O & $s_{1/2}$ & 0.0 & 3.6 &   2.177 & 1.67 & 1.28  \\

$^{40}$Ca & $^{39}$K &  $d_{3/2}$ & 0.0  & 8.33  &   4.208 & 3.12 & 2.63 & 2.58(19) \\
         &          &   $s_{1/2}$ & 2.52 & 10.85 &   2.104 & 1.53 & 1.12 & 1.03(7)  \\
$^{40}$Ca & $^{39}$Ca & $d_{3/2}$ & 0.0  & 15.64 &   4.208 & 3.06 & 2.54 \\
         &          &   $s_{1/2}$ & 2.47 & 18.11 &   2.104 & 1.58 & 1.05  \\        
$^{48}$Ca & $^{47}$K & $s_{1/2}$ & 0.0 & 15.81 &   2.086 & 1.68 & 1.18 & 1.07(7) \\
         &          & $d_{3/2}$ & 0.36 & 16.17 &   4.172 & 3.18 & 2.64 & 2.26(16)\\
%         &          & $d_{5/2}$ & 3.43 & 19.24 &  6.258 & 4.21 \\
$^{48}$Ca & $^{47}$Ca & $f_{7/2}$ & 0.0 & 9.95 &  8.523 & 5.24 & 4.79 \\
$^{56}$Ni & $^{55}$Co & $f_{7/2}$ & 0.0 & 7.17 &   8.444 & 5.44 & 5.00 \\
$^{56}$Ni & $^{55}$Ni & $f_{7/2}$ & 0.0 & 16.64 &   8.444 & 5.31 & 4.83 \\
$^{132}$Sn & $^{131}$In & $g_{9/2}$ & 0.0 & 15.71 &   8.247 & 6.76 & 6.16 \\
%         &          & $p_{1/2}$ & 0.30 & 16.01 &   2.0 &  1.47  \\
%         &          & $p_{3/2}$ & 1.29 & 17.00 &   4.0 &  2.94 \\
$^{132}$Sn & $^{131}$Sn & $d_{3/2}$ & 0.0 & 7.31 &   4.0 & 3.73 & 2.52 \\
%         &          & $h_{11/2}$ & 0.07 & 7.38 &   12.47 & 6.45 \\
%         &          & $s_{1/2}$ & 0.33 & 7.31 &   2.0 & 1.68  \\
%         &          & $d_{5/2}$ & 1.66 & 8.97 &   6.0 & 4.88  \\
%         &          & $g_{7/2}$ & 2.43 & 9.75 &   8.0 & 6.10  \\  
$^{208}$Pb & $^{207}$Tl & $s_{1/2}$ & 0.0 & 8.00  &   2.0 & 1.64 & 1.04 & 0.98(9)\\
           &            & $d_{3/2}$ & 0.35 & 8.35 &   4.0 & 3.28 & 2.52 & 2.31(22)\\
           &            & $h_{11/2}$ & 1.35 & 9.35 &   12. & 7.24 & 6.70 & 6.85(68)\\
           &            & $d_{5/2}$ & 1.67 & 9.67 &   6.0 & 5.00 & 4.22 & 2.93(28) \\
           &            & $g_{7/2}$ & 3.47 & 11.47 &  8.0 & 6.36 & 5.55 & 2.06(20)\\
$^{208}$Pb & $^{207}$Pb & $p_{1/2}$ & 0.0 & 7.37  &   2.0 & 1.79 & 0.83 \\
\hline
\hline 
\end{tabular}
 
\end{center}
\label{table1}
\end{table} 

\begin{figure}[t!]
\centering 
\vspace{1.0 cm}
\includegraphics[scale=0.4]{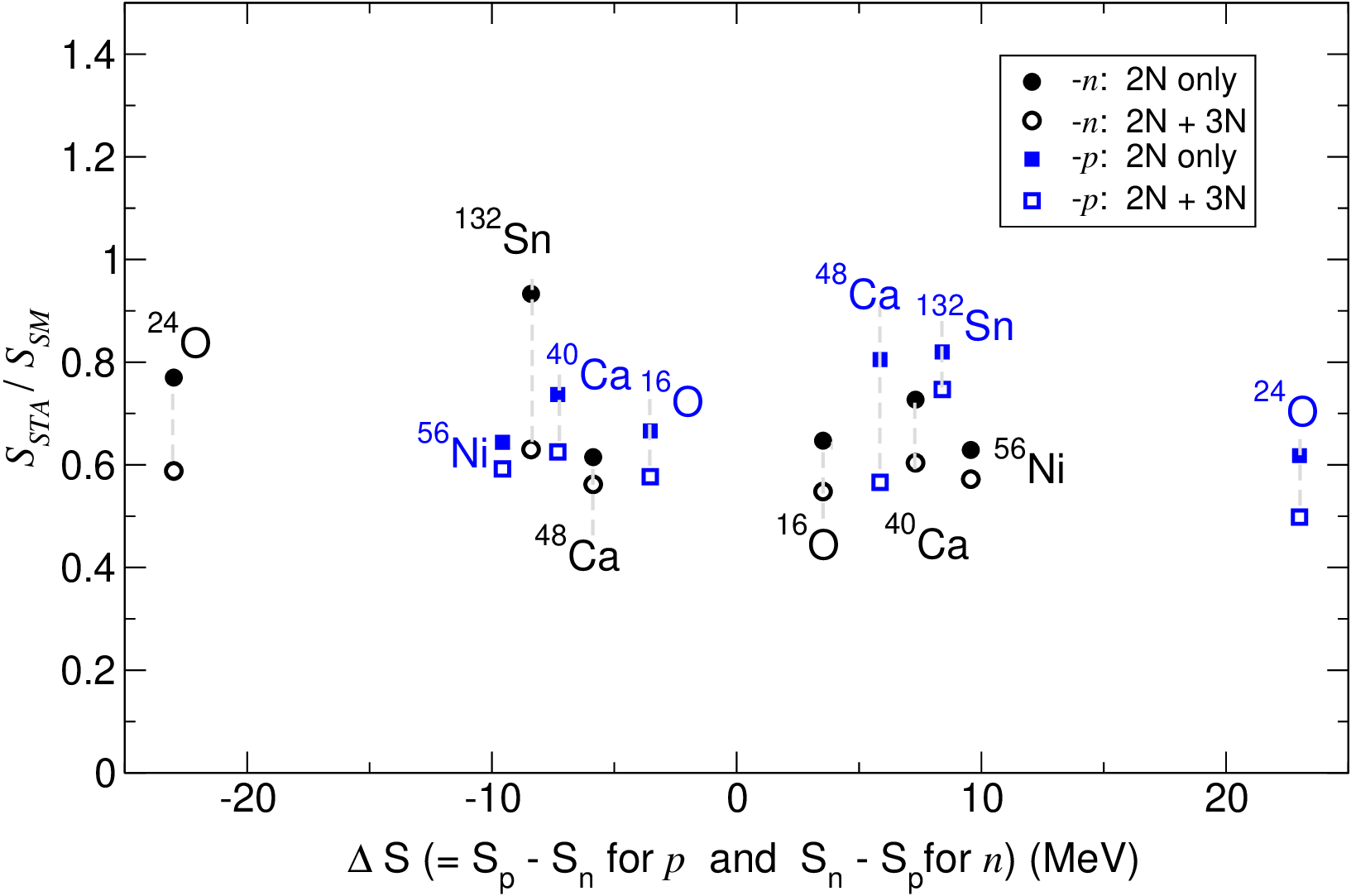}
\caption{The ratio $S_{\rm STA}/S_{\rm SM}$ for neutron and proton removals from $^{16,24}$O, $^{40,48}$Ca, $^{56}$Ni   and $^{132}$Sn, calculated with NN force only and with including the 3N force having   $W_0= 1$ MeV and $\rho_0 = 2$ fm. The ratios are shown as functions of difference $\Delta S$ between the proton ($S_p$) and neutron ($S_n$) separation energies. 
}
\label{fig:reduction}
\end{figure}

The  spectroscopic factors corresponding to the overlaps with NN force only and with added spin- and isospin-independent 3N contribution  $W^{(0)}$ are shown in Table 1. For proton removal from $^{16}$O, $^{40,48}$Ca and $^{208}$Pb, the spectroscopic factors either agree or are very close to those extracted from the $(e,e'p)$ study, except for removing $g_{7/2}$ proton from $^{208}$Pb. They are reduced with respect to the $0\hbar\omega$ shell model values (see Table 1), which for double-magic nuclei are the same as those given by the independent particle model.  To investigate if there is any asymmetry in the spectroscopic factor reductions the overlap functions for neutron removal were also calculated  for the same targets.  Three more cases involving unstable nuclei, $^{24}$O, $^{56}$Ni and $^{132}$Sn, where the difference between the proton and neutron separation energies is larger, were added to this study. The oscillator radii for these nuclei were chosen to be 1.85, 1.92 and 2.15 fm, respectively. The ratio  of the STA spectroscopic factors to those obtained in the $0\hbar\omega$ shell model is shown in Fig. \ref{fig:reduction} as a function of the difference between the proton and neutron (or neutron and proton for neutron removal) separation energies  $\Delta S$. For $^{16}$O, $^{40}$Ca and $^{56}$Ni there is practically no asymmetry both with and without the 3N force. Strong and unusual asymmetry predicted for $^{48}$Ca with NN force only disappears when the 3N force is added. For the case of $^{24}$O,  with the largest $\Delta S$,  the asymmetry   seen in $S_{\rm STA}/S_{\rm SM}$ calculated with NN force only also becomes smaller when 3N force is added. The case of $^{132}$Sn is very unusual. Without 3N force, the reduction factor $S_{\rm STA}/S_{\rm SM} $ is 0.93 and 0.83 for neutron and a more strongly bound proton, respectively.
 However, the contribution from the 3N force is larger for neutron than for proton removal so that adding 3N force makes neutron spectroscopic strength reduction stronger than the one for  proton. This conflicts with experimental observations made with knockout reactions in \cite{Vaq20} suggesting that a different choice of 3N should take place. Indeed, with the second single-term 3N force choice mentioned above, $W^{(\sigma \tau)}_0=-0.9$ MeV and $\rho_0= 2$ fm, the  $S_{\rm STA}/S_{\rm SM}$ ratio becomes very similar for neutron and proton, being 0.73 and 0.76, respectively, suggesting no asymmetry in the spectroscopic strength reduction. 
   Not shown in Fig. \ref{fig:reduction} is the $^{208}$Pb case where, as in the case of $^{132}$Sn,  the 3N contribution is much stronger than that in the proton case leading to the
$S_{\rm STA}/S_{\rm SM}$ value of 0.41 which at first sight   seems to be unrealistic.  With the second choice of the 3N force, $W^{(\sigma \tau)}_0=-0.9$ MeV and $\rho_0= 2$ fm, the spectroscopic factor is 1.08 and 1.09 for neutron and proton, respectively, with the same ratio  $S_{\rm STA}/S_{\rm SM}=0.54$ for both.

The ANCs and r.m.s. radii for the overlaps from Table 1 are shown in Table 2. One can see that including the 3N force gives smaller values of the ANC $C_{lj}$ and larger radii. The single-particle ANCs $b_{lj} = C_{lj}/\sqrt{S_{lj}}$ can be either smaller of larger than those obtained without 3N force. The coefficients $b_{lj}$ give an idea about the width of the effective potential well that would be needed to generate the overlaps in a two-body model. Including 3N force does not have a unique effect on this width. There is not much information on experimental ANCs for the nuclei considered here, apart from the proton removal from $^{16}$O. The $(^3He,d)$ reaction study in \cite{Art03}
gives $C^2_{lj} = 298(63)$ fm$^{-1}$ for the ground state of $^{15}$N while a corrected value of this ANC from Ref. \cite{Muk08}, based on the updated value of the ANC for $^3$He, is $C^2_{lj} = 175(29)$ fm$^{-1}$. The latter covers both the NN and NN+3N values of the ANC obtained in the STA. Also, the r.m.s. radius for the $\la^{15}$N$|^{16}$O$\ra$ overlap, calculated with NN+3N force, agrees with the experimental values 2.943(30) and 2.719(24) fm for $\frac{1}{2}^-$ and $\frac{3}{2}^-$ states, respectively, obtained in \cite{Leu94}.

\begin{table}[h]
\caption{ The ANC $C_{lj}$, single-particle ANC $b_{lj}$ (both in fm$^{-1/2}$) and the r.m.s. radius (in fm) of the overlap functions $\la A | A-1\ra $ obtained  in STA with NN force only and  with including 3N force. 
 The quantum numbers $lj$ of the removed nucleon are shown in the third column and eN denotes 10$^N$.
 } 
\begin {center}
\begin{tabular}{ p{0.8cm} p{1.0 cm} p{1.0 cm} p{1.5 cm} p{1.5 cm} p{1.7 cm}
p{1.5 cm} p{1.7 cm} p{1.0 cm}}
%\begin{tabular}{ ccccccccccc}
\hline
\hline
 & & & \multicolumn{3}{c}{NN} & \multicolumn{3}{c}{NN + 3N}\\
 \cline{4-6} \cline{7-9}
 $A$ &  $A$$-$1 & $lj$ & $C_{lj}$ & $b_{lj}$ & 
   $\la r^2\ra^{1/2}$  & $C_{lj}$ & $b_{lj}$ & 
   $\la r^2\ra^{1/2}$  
 \\
 \hline
$^{ 16}$O     & $^{ 15}$N     & $p_{ 1/2}$   &     13.4     &     11.2     &   2.871 &     12.9     &     11.6     &   2.905\\
      &      & $p_{ 3/2}$   &     35.6     &     22.2     &   2.774 &     33.1     &     21.4     &   2.791\\
 $^{ 16}$O     & $^{ 15}$O     & $p_{ 1/2}$   &     10.7     &     9.11     &   2.823 &     10.2     &     9.43     &   2.866\\
      &      & $p_{ 3/2}$   &     26.8     &     16.8     &   2.744 &     26.3     &     17.2     &   2.767\\
 $^{ 24}$O     & $^{ 23}$N     & $p_{ 1/2}$   &     70.4     &     62.0     &   2.830 &     67.8     &     66.5     &   2.903\\
 $^{ 24}$O     & $^{ 23}$O     & $s_{ 1/2}$   &    -3.51     &    -2.72     &   3.400 &    -3.15     &    -2.78     &   3.445\\
 $^{ 40}$Ca    & $^{ 39}$K     & $d_{ 3/2}$   &     57.8     &     32.7     &   3.554 &     55.6     &     34.3     &   3.616\\
     &     & $s_{ 1/2}$   &    -132.     &    -107.     &   3.392 &    -123.     &    -116.     &   3.527\\
 $^{ 40}$Ca    & $^{ 39}$Ca    & $d_{ 3/2}$   &     22.8     &     13.0     &   3.482 &     21.9     &     13.7     &   3.551\\
    &    & $s_{ 1/2}$   &    -48.6     &    -38.7     &   3.131 &    -45.1     &    -44.0     &   3.380\\
 $^{ 48}$Ca    & $^{ 47}$K     & $s_{ 1/2}$   &    -258.     &    -199.     &   3.501 &    -242.     &    -222.     &   3.695\\
     &      & $d_{ 3/2}$   &     162.     &     90.7     &   3.493 &     157.     &     96.4     &   3.567\\
 $^{ 48}$Ca    & $^{ 47}$Ca    & $f_{ 7/2}$   &     7.69     &     3.36     &   3.934 &     7.53     &     3.44     &   3.966\\
 $^{ 56}$Ni    & $^{ 55}$Co    & $f_{ 7/2}$   &     146.     &     62.7     &   3.978 &     144.     &     64.3     &   4.008\\
 $^{ 56}$Ni    & $^{ 55}$Ni    & $f_{ 7/2}$   &     31.8     &     13.8     &   3.909 &     31.3     &     14.2     &   3.948\\
 $^{132}$Sn    & $^{131}$In    & $g_{ 9/2}$   &    1.32e4 &    5.08e3 &   4.875 &    1.30e4 &    5.24e3 &   4.927\\
 $^{132}$Sn    & $^{131}$Sn    & $d_{ 3/2}$   &    -24.5     &    -12.3     &   4.867 &    -22.3     &    -13.5     &   5.218\\
 $^{208}$Pb    & $^{207}$Tl    & $s_{ 1/2}$   &    1.21e7 &    9.45e6 &   5.488 &    1.15e7 &    1.13e7 &   5.943\\
     &     & $d_{ 3/2}$   &   -9.33e6 &   -5.15e6 &   5.507 &   -9.21e6 &   -5.80e6 &   5.785\\
     &     & $h_{11/2}$   &    1.93e6 &    7.17e5 &   5.867 &    1.92e6 &    7.42e5 &   5.919\\
     &    & $d_{ 5/2}$   &   -8.24e6 &   -3.69e6 &   5.481 &   -8.11e6 &   -3.95e6 &   5.674\\
     &   & $g_{ 7/2}$   &    2.14e6 &    8.49e5 &   5.486 &    2.12e6 &    9.00e5 &   5.579\\
 $^{208}$Pb    & $^{207}$Pb    & $p_{ 1/2}$   &     47.6     &     35.6     &   5.176 &     41.6     &     45.7     &   6.094\\

\hline
\hline 
\end{tabular}
 
\end{center}
\label{table2}
\end{table}

\section{Summary and future perspectives}

In this work, an effective  3N force has been introduced into the source term approach designed to generate one-nucleon overlap functions for various nucleon-removal reactions. This force acts between the  removed nucleon and two nucleons of the residual nucleus and it arises because the STA  uses nuclear wave functions calculated in a truncated model space, more specifically, the one given by the 0$\hbar \omega$ shell model.  The 3N force in STA is not equal to the 3N force that could be  employed  in modelling many-body nuclear wave functions in exactly  the same way as the 2N force between the removed nucleon and one nucleon in the residual nucleus in STA is not the same as the effective 2N force employed in the wave function calculations \cite{Tim09,Tim10}.
Although, in principle, an STA-rated induced 3N force could be calculated in a microscopic approach such a task would require a major effort. Therefore, this paper adopts a phenomenological approach in choosing such a force assuming that it has four components determined by a different spin and isospin content and has a form given by a hypercentral gaussian potential.

 The first application of the effective 3N force have been made here for one-proton removal from double-magic nuclei $^{16}$O, $^{40,48}$Ca and $^{208}$Pb where phenomenological overlap functions are available from the $(e,e'p)$ study. The calculations revealed that only one component, spin- and isospin-independent, with the hypercentral range of 2 fm and a depth of 1 MeV is sufficient to improve description of these phenomenological overlaps. The corresponding spectroscopic factors either agree with or are very close to the $(e,e'p)$ values. These spectroscopic factors are reduced with respect to those given by the independent particle model, or the $0\hbar \omega$ shell model. Reduction is also predicted for one-neutron removal from the same double-magic nuclei and for $^{16}$O and $^{40}$Ca it is in the same proportions both for neutrons and protons irrespective of the presence of the 3N force, which slightly decreases these spectroscopic factors. However, for $^{48}$Ca a large asymmetry in reduction is seen with the 2N force only, which is removed when 3N force is included. Including spin- and isospin-independent 3N force leads to unusual $S_{\rm STA}/S_{\rm SM}$ results for heavier nuclei $^{132}$Sn and $^{208}$Pb. A different choice of the 3N force, represented by a forth term in Eq. (\ref{3Nmodel}), suggests no asymmetry in spectroscopic factor reduction in these nuclei.

 Whether the spectroscopic factor reduction in  STA can be reliably studied for an  arbitrary nucleus  depends on the existence of a universal effective 3N force applicable to a wide range of atomic nuclei. No attempts to find such a force have been made here  because of several reasons. 1) Better quality data on ``experimental" overlap functions are needed. Currently, available phenomenological overlaps have large experimental and systematic uncertainties. However, reducing these uncertainties is a challenging task because the overlap functions are not observables and they can only be indirectly deduced from reaction data analysis.  2)  The STA overlaps depend,   in the first instance, on the effective 2N force and harmonic oscillator radius. Any attempts of  fitting the 3N force should be accompanied by fitting 2N force as well. At the moment, in all previous publications the 2N force was fixed. However, it could be possible to find a better representation of this force, in particular, it could depend on oscillator radius as well so that the resulting overlaps would not be subjected to strong variations with this radius. 3) The hypercentral gaussian functional form used in this work for effective 3N force may not represent it in the most optimal way. Other options, such as a symmetrized product of two NN formfactors should also be  explored together with other structures such as those mimicking double-pion exchange. Future work will tackle these issues.
 
 Finally, going beyond the $0\hbar \omega$ approximation for nuclear models on $A$ and $A-1$ can be very important for the further development of the STA. It will allow  the wave functions to be used which describe better the internal nuclear region, in particular, it can also reduce the dependence of the STA overlaps on the oscillator radius choices. While it could be difficult to do it for medium-mass nuclei, for light nuclei this is certainly possible. Having a good STA-rated 2N and 3N effective interactions consistent with extended model space will make it possible to predict the overlap functions between any nuclear states accessible to extended shell model studies.

\section*{Acknowledgements}
This work was supported by the United Kingdom Science and Technology Facilities Council (STFC) under Grant No.  ST/P005314/1. An early stage of this work enjoyed some help from A. Matta and M. Moukaddam, which is greatly appreciated.

\appendix
\section{STA matrix elements of the 3N interaction.}

\subsection{Spatial part}
We will first derive an expression to evaluate the spacial part 
\beq
\fl
\la \varphi_{n'_1l'_1m'_1}(\ve{r}_1) \varphi_{n'_2l'_2m'_2}(\ve{r}_2) e^{i \ve{k}\ve{r}_3} \mid 
W_{123}
\mid
\varphi_{n_1l_1m_1}(\ve{r}_1)\varphi_{n_2l_2m_2}(\ve{r}_2)\varphi_{n_3l_3m_3}(\ve{r}_3)\ra
\eeqn{basicME}
of the matrix element (\ref{indME}) assuming that the 3N force $W_{123}$ has a hypercentral form given by Eq. (\ref{WHC}). The matrix element (\ref{indME}) is evaluated in the harmonic oscillator basis defined by the single-particle wave functions $\varphi_{nlm}(\ve{r}) = \varphi_{nl}(r) Y_{lm}(\hat{\ve{r}})$ with the  radial part   
\beq
\varphi_{nl}(r) =\frac{1}{b^{3/2}} \sqrt{\frac{2n!}{\Gamma \left(n+l+\frac{3}{2}\right)} }\left(\frac{r}{b}\right)^l e^{-\frac{r^2}{2b^2}}L_n^{l+1/2} \left(\frac{r^2}{b^2}\right),
\eeqn{HOwf}
where $b$ is the oscillator length. We will also need harmonic oscillator wave functions
\beq
\phi_{nlm}(\ve{q}) = (-)^n i^l(2\pi)^{3/2}  \phi_{nl}(q) Y_{lm}(\hat{\ve{q}})
\eeqn{wfq}
in momentum space with $\phi_{nl}(q)$ given by Eq. (\ref{HOwf}) in which the substitution $b \rightarrow 1/b$ is made.  As a first step to evaluate (\ref{basicME}) we apply relation
\beq
\fl
\varphi^*_{n'l'm'}(\ve{r}) \varphi_{nlm}(\ve{r}) =  \sum_{\nu NLM } (-)^{m'}(l'-m' lm | LM) \la \nu 0 NL:L | n'l' nl:L\ra 
\eol \,\,\,\,\,\,\,\,\,\,\,\,\,\,\,\, \,\,\,\,\,\,\,\,\,\,\,\,\,\,\,\, \,\,\,\,\,\,\,\,\,\,\,\,\,\,\,\times
\varphi_{\nu 0 0}(0) \varphi_{NLM}(\sqrt{2} \ve{r})
\eeqn{HOprod}
to the products of wave functions of the same coordinates, which are either $\ve{r}_1$ or  $\ve{r}_2$. In Eq. (\ref{HOprod}) the $\la \nu 0 NL:L | n'l' nl:L\ra$ is the Moshinsky bracket for particle with equal masses \cite{Moshinsky}. Introducing new variables, the normalized Jacobi coordinates $\ve{\xi}_1 = (\ve{r}_1 - \ve{r}_2)/\sqrt{2}$ and $\ve{\xi}_2 = \sqrt{\frac{2}{3}}\left( \frac{1}{\sqrt{2}}\ve{X}_{12}-\ve{r}_3\right)$, and applying the Moshinsky transformation again to the function of $\ve{r}_1$ and $\ve{r}_2$we get
\beq
\fl
\la \varphi_{n'_1l'_1m'_1}(\ve{r}_1) \varphi_{n'_2l'_2m'_2}(\ve{r}_2) e^{i \ve{k}\ve{r}_3} 
\mid 
W_{123}
\mid
\varphi_{n_1l_1m_1}(\ve{r}_1)\varphi_{n_2l_2m_2}(\ve{r}_2)\varphi_{n_3l_3m_3}(\ve{r}_3)\ra \eol \fl
= 
%\sum_{L_1L_2L_3} 
\sum \,\, (-)^{m'_1} (l'_1 -m'_1 l_1 m_1 | L_1 M_1) \,
 \la \nu_1 0 N_1 L_1: L_1| n'_1l'_1n_1l_1:L_1\ra  \, \varphi_{\nu_1 0 0}(0)
\eol \fl \,\,\,\,\,\,\,\,\,\,\,\,\,\,\,\,\,
\times (-)^{m'_2} (l'_2 -m'_2 l_2 m_2 | L_2 M_2) \,
 \la \nu_2 0 N_2 L_2: L_2| n'_2l'_2n_2l_2:L_2\ra  \,\varphi_{\nu_2 0 0}(0)
 \eol  \fl \,\,\,\,\,\,\,\,\,\,\,\,\,\,\,\,\, \times
(L_1M_1L_2M_2| L_{12}M_{12})(\lambda \mu \,L_3 M_3 | L_{12}M_{12}) 
\la \nu \lambda N'_3 L_3 :L_{12} | N_1L_1N_2L_2:L_{12}\ra 
\eol \fl \times
{ W}_0 \int d\ve{\xi}_1 d\ve{X}_{12} d\ve{r}_3 \, e^{-i   \ve{k}\ve{r}_3}
 e^{-\frac{\xi_1^2+\xi_2^2}{\rho_0^2}}
\varphi_{\nu\lambda \mu}(\sqrt{2} \ve{\xi}_1) \varphi_{N'_3 L_3M_3}(\sqrt{2}\ve{X}_{12})  \varphi_{n_3l_3m_3}(\ve{r}_3),
\eeqn{ME1}
where the sum runs over $\{\nu_1N_1L_1\nu_2N_2L_2 \nu \lambda {\tilde n}'_3 l'_3 \}$ and $\{M_1M_2M_3\mu m'_3\}$. We proceed with integrating over $d\ve{\xi}_1$ using
\beq
\int d\ve{\xi}_1 e^{-\frac{\xi_1^2}{\rho_0^2}}\varphi_{\nu\lambda \mu}(\sqrt{2} \ve{\xi}_1)
 =  w_{\nu}\delta_{\lambda,0} \delta_{\mu,0}.
\eeqn{x1}
Then we aim to make an integration over $d\ve{X}_{12}$. For this purpose we
note that $e^{-\xi^2_2/\rho_0^2} = \sqrt{2\pi\Gamma(\frac{3}{2})} \varphi_{000}\left(\frac{\sqrt{2}b}{\rho_0} \ve{\xi}_2\right)$. Introducing another variable change, $\ve{y}_1 =\alpha \sqrt{2} \ve{X}_{12} - \beta \frac{\sqrt{2}b}{\rho_0} \ve{\xi}_2 $ and  $\ve{y}_2 =\beta \sqrt{2} \ve{X}_{12} + \alpha \frac{\sqrt{2}b}{\rho_0} \ve{\xi}_2 $ with $\alpha = (1+\frac{3\rho_0^2}{b^2})^{-1/2}$ and $\beta =(1+\frac{b^2}{3\rho_0^2})^{-1/2}$, we rearrange the product $e^{-\xi^2_2/\rho_0^2}
 \varphi_{N'_3 L_3M_3}(\sqrt{2}\ve{X}_{12})$ using Moshinsky technique for particle with different masses \cite{Smirnov}
\beq
\fl
|  \varphi_{N'_3 L_3}(\sqrt{2} \ve{X}_{12})\varphi_{00} \left(\frac{\sqrt{2}b}{\rho_0}\ve{\xi}_2\right):L_3M_3 \ra  
\eol
\fl
= \sum_{N_3N} \la N_3L_3 NL:L_3|d_1|  N'_3 L_3 00:L_3 \ra 
| \varphi_{N_3L_3}(2\alpha \ve{r}_3) \varphi_{NL}(\ve{y}_2):L_3M_3\ra,
\eeqn{erase1}
where $d_1 = \frac{b^2}{3\rho_0^2}$. The numerical values of the Moshinsky brackets $\la N_3L_3 NL:L_3|d_1| 00 N'_3 L_3:L_3 \ra $ were calculated using the formalism and the Fortran code from \cite{Trlifaj}.
%\beq
%c_N = \int d\ve{r} \varphi_{N00} (\ve{r}) =
%\eeqn{cNdef}
After integration over $d\ve{X}_{12}$  we integrate over $d\ve{r}_3$ and obtain
\beq
\fl
\int d\ve{r}_3 \, \varphi_{N_3L_3 M_3 }(2\alpha\ve{r}_3) e^{- i\ve{k}\ve{r}_3} \varphi_{n_3 l_3 m_3 }(\ve{r}_3)   =  (-)^{l'_3}\,\theta^3
\sum_{\nu_3 N_3 L_3 M_3n'_3 l'_3 m'_3}
 (L_3 M_3 l_3 m_3 | l'_3 m'_3) \eol \times
 \la \nu_3 0 n'_3 l'_3: l'_3| d_2| N_3L_3n_3l_3:l'_3\ra 
 \varphi_{\nu_3 0 0}(0) \phi_{n'_3l'_3m'_3}\left(\theta  \ve{k}\right),
\eeqn{fef}
where $d_2 = (4\alpha^2)^{-1}$,  $\theta = (1+4\alpha^2)^{-1/2}$. This leads to the final result for the matrix element (\ref{ME1}),
\beq
\fl
\la \varphi_{n'_1l'_1m'_1}(\ve{r}_1) \varphi_{n'_2l'_2m'_2}(\ve{r}_2) e^{i \ve{k}\ve{r}_3} \mid 
W_{123}
\mid
\varphi_{n_1l_1m_1}(\ve{r}_1)\varphi_{n_2l_2m_2}(\ve{r}_2)\varphi_{n_3l_3m_3}(\ve{r}_3)\ra \eol \fl
= 
\sum \,\, (-)^{m'_1+m'_2+m'_3} (l'_1 -m'_1 l_1 m_1 | L_1 M_1) \,
  (l'_2 -m'_2 l_2 m_2 | L_2 M_2) \,
(l'_3 -m'_3  l_3 m_3 | L_{3}M_{3}) \,
 \eol  \fl \,\,\,\,\,\,\,\,\,\,\,\,\,\,\,\,\,  
\times (-)^{L_3+M_3} \hat{L}_3^{-1}  (L_1M_1L_2M_2| L_{3}-M_{3}) \,
W_{\gamma_1\gamma_2\gamma_3}\, \phi_{n'_3 l'_3 m'_3}\left(\theta  \ve{k}\right),
\eeqn{ME1final}
where $\gamma_i = \{n'_il'_in_il_iL_i\}$, $i = 1,2,3$ and
\beq
\fl
W_{\gamma_1\gamma_2\gamma_3} =
 \left( \frac{\sqrt{\pi} }{2}\, b\beta^2  \theta^2\right)^{\frac{3}{2}} \, \hat{l}'_3 { W}_0   
\sum_{N_1N_2N_3N'_3 N\nu_1\nu_2\nu_3\nu}  w_{\nu}
 \phi_{N00}(0) \varphi_{\nu_1 0 0}(0)\varphi_{\nu_2 0 0}(0)\varphi_{\nu_3 0 0}(0)
 \eol \fl
 \times
 \la \nu_1 0 N_1 L_1{\rm :}L_1| n'_1l'_1n_1l_1{\rm :}L_1\ra  \, 
   \la \nu_2 0 N_2 L_2{\rm :}L_2| n'_2l'_2n_2l_2{\rm :}L_2\ra  \, \la \nu 0 N'_{3}L_{3} {\rm :}L_{3} | N_1L_1N_2L_2{\rm :}L_{3}\ra
  \,
 \eol  \times
 \la N_{3}L_{3} N0{\rm :}L_{3} | d_1|  N'_{3}L_{3} 00{\rm :}L_{3}\ra 
  \la \nu_3 0 n'_3 l'_3{\rm :}l'_3|d_2| N_{3}L_{3}n_3l_3{\rm :}l'_3\ra.
\eeqn{Wnls}
The summation in (\ref{ME1final}) is performed over $\{  n'_3 l'_3m'_3 L_1  L_2  L_3 \}$ while $M_i = m_i - m'_i$ for all $i = 1,2,3$. It should be noted that (\ref{ME1final}) is in fact an expansion over a finite number of single-particle harmonic oscillator wave functions   $\phi_{n'_3l'_3 m'_3}$ of  an argument $\theta \ve{k} $ that depends on the range of the 3N interaction. In Eq. (\ref{ME1final}) $2n'_3+l'_3$ changes from 0 to $2n_1+l_1+2n_2+l_2+2n_3+l_3+2n'_1+l'_1+2n'_2+l'_2$.
In principle, the matrix element (\ref{ME1}) could be sought for as an expansion over the basis $\phi_{n'_3 l'_3 m'_3}(\ve{q})$ that does not depend on the choice of the 3N force. However, such an expansion would need an infinite sum over $n'_3 l'_3$ so that convergence issue could slow the calculations.

\subsection{Adding spin and isospin variables}

To calculate the source term we will use the single-particle wave functions in the $j$-$j$ coupling:
\beq
\varphi_{\alpha_j}(\ve{r}) =\sum_{m\sigma} (lm \dem \sigma | j m_j) \varphi_{nlm}(\ve{r}) \chi_{\sigma},
\eeqn{spwfjj}
where $\alpha_i \equiv \{ n_il_ij_im_{j_i}\}$ and $\chi_{\sigma}$ is the spin function of the nucleon.  To calculate the matrix element
\beq
\la \varphi_{\alpha'_1}(\ve{r}_1) \varphi_{\alpha'_2}(\ve{r}_2) e^{i \ve{k}\ve{r}_3} 
\mid 
W_{123}
\mid
\varphi_{\alpha_1}(\ve{r}_1)\varphi_{\alpha_2}(\ve{r}_2)\varphi_{\alpha_3}(\ve{r}_3)\ra 
\eeqn{MEjj}
we need results (\ref{ME1final}) and (\ref{Wnls}) from previous subsection and
the following summations over projections of angular momenta:
\beq
\sum_{m_1m'_1\sigma_1} (-)^{m'_1} (l'_1 -m'_1 l_1 m_1 | L_1 M_1) \, (l'_1m'_1 \dem \sigma'_1 | j'_1 m'_{j_1})(l_1m_1 \dem \sigma_1 | j_1 m_{j_1})
\eol
=(-)^{l_1+j_1 + m'_{j_1}}  \hat{j}_1 \hat{j}'_1 (j'_1-m'_{J_1} j_1 m_{j_1} |L_1 M_1) W(L_1l_1j'_1\dem;l'_1 j_1)
\eeqn{x2}
and
\beq
\fl 
\sum_{m_3m'_3 \sigma_3}
(-)^{m_3} (l'_3 -m'_3 l_3 m_3|L_3-M_3) (l_3 m_3\dem \sigma_3 | j_3 m_{j_3})\phi_{n'_3 l'_3m'_3} (\theta \ve{k})\chi_{\sigma_3} \eol
\fl
=
\sum_{j'_3 m'_{j_3}} (-)^{l_3+j_3 + m_{j_3}}  \hat{j}_3
\hat{j}'_3 (j'_3 -m'_{j_3} j_3 m_{j_3} |L_3 -M_3) W(L_3l_3j'_3\dem;l'_3 j_3)
\phi_{n'_3 l'_3j'_3m'_{j_3}}(\theta \ve{k}).
\eeqn{summ3}
It gives  
\beq
%\fl
\la \varphi_{\alpha'_1}(\ve{r}_1) \varphi_{\alpha'_2}(\ve{r}_2) e^{i \ve{k}\ve{r}_3} 
\mid 
W_{123}
\mid
\varphi_{\alpha_1}(\ve{r}_1)\varphi_{\alpha_2}(\ve{r}_2)\varphi_{\alpha_3}(\ve{r}_3)\ra \eol 
%\fl
= 
%\sum_{L_1L_2L_3} 
\sum_{\alpha'_3} %{n'_3 l'_3 j'_3 m'_{j_3}}   
%\phi_{n'_3 l'_3 j'_3 m'_{j_3}}
\phi_{\alpha'_3}
\left(\theta  \ve{k}\right)
\sum_{L_1L_2 L_3}  (-)^{L_3+M_3}  \hat{L}_3^{-1}(L_1M_1L_2M_2| L_3 -M_3)
W_{\gamma_1\gamma_2\gamma_3}
%W_{n'_1l'_1n'_2l'_2n'_3l'_3 n_1l_1n_2l_2n_3l_3}^{L_1L_2 L_3}
\eol %\fl \,\,\,\,\,\,\,\,\,\,\,\,\,\,\,\,\,
\times
(-)^{l_1 - j_1 -m'_{j_1}} \, \hat{j}_1 \hat{j}'_1(j'_1 -m'_{j_1} j_1 m_{j_1} | L_1 M_1) \,
W(L_1 l_1j'_1 \dem ; l'_1 j_1)
%\eol \fl \,\,\,\,\,\,\,\,\,\,\,\,\,\,\,\,\,
\eol 
\times (-)^{l_2 - j'_2 -m'_{j_2}} \, \hat{j}_2 \hat{j}'_2 (j'_2 -m'_{j_2} j_2 m_{j_2} | L_2 M_2) \,
W(L_2 l_2j'_2 \dem ; l'_2 j_2)
\eol %\fl \,\,\,\,\,\,\,\,\,\,\,\,\,\,\,\,\,
\times (-)^{l_3 - j'_3 -m'_{j_3}} \, \hat{j}_3 \hat{j}'_3 (j'_3 -m'_{j_3} j_3 m_{j_3} | L_3 M_3) \,
W(L_3 l_3j'_3 \dem ; l'_3 j_3),
\eeqn{ME2final}
where  $W_{\gamma_1\gamma_2\gamma_3}$ is given by Eq. (\ref{Wnls}). This resut was obtained using $(-)^{m'_{j_1}+m'_{j_2}+m_{j_3}} = (-)^{ m'_{j_1}+m'_{j_2}+m'_{j_3}+M_3}$  and $(-)^{l_3} = (-)^{L_3+l'_3}$.

If nucleons carry isospin quantum numbers $\tau_i$ then the corresponding matrix elements of the hypercentral 3N force are expressed by Eq. (\ref{ME2final}) supplemented by $\delta_{\tau'_1\tau_1}\delta_{\tau'_2\tau_2}\chi_{\tau_3}$.

\subsection{3N force containing the $(\ve{\tau}_i \cdot \ve{\tau}_j)$ terms.}

For hypercentral 3N force considered in this paper
\beq
\fl
W_{123}^{(\tau)} \equiv
\sum_{i\neq k<j \neq k} (\ve{\tau}_i \cdot \ve{\tau}_j) W_{k,ij}^{(\tau)} 
= \left[
(\ve{\tau}_1 \cdot \ve{\tau}_2)+(\ve{\tau}_1 \cdot \ve{\tau}_3)+(\ve{\tau}_2 \cdot \ve{\tau}_3)\right] W_{123}^{(\tau)}(\rho_{123}) 
\eeqn{Wtau}
so that the matrix element
$
\la \varphi_{\alpha'_1}(\ve{r}_1) \varphi_{\alpha'_2}(\ve{r}_2) e^{i \ve{k}\ve{r}_3} \chi_{\tau'_3}
\mid 
W^{(\tau)}_{123}
\mid
\varphi_{\alpha_1}(\ve{r}_1)\varphi_{\alpha_2}(\ve{r}_2)\varphi_{\alpha_3}(\ve{r}_3)\ra $, where $\alpha_i$ now includes the isospin projection variable $\tau_i$ as well, is equal to  the matrix element given by Eq. (\ref{ME2final}) times
\beq
\Xi = \la \chi_{\tau'_1}\chi_{\tau'_2}\chi_{\tau'_3} \mid
(\ve{\tau}_1 \cdot \ve{\tau}_2)+(\ve{\tau}_1 \cdot \ve{\tau}_3)+(\ve{\tau}_2 \cdot \ve{\tau}_3 )\mid 
 \chi_{\tau_1}\chi_{\tau_2}\chi_{\tau_3} \ra
 \eeqn{taupart}
Using
 \beq
 \fl
 \la \chi_{\tau'_i}\chi_{\tau'_j} |
\ve{\tau}_i \cdot \ve{\tau}_j |
 \chi_{\tau_i}\chi_{\tau_j}  \ra  \equiv X_{\tau'_i \tau'_j \tau_i \tau_j} =
  2 \sum_t (-)^{\tau_i+\tau_j +1+t}( \dem \tau'_i \,\,\dem -\tau_i \mid 1 -t)
(\dem \tau'_j \,\,\dem -\tau_j \mid 1 t)
\eol
\eeqn{tautau}
we obtain 
\beq
\Xi = \delta_{\tau'_1 \tau_1} X_{\tau'_2 \tau'_3 \tau_2 \tau_3}+\delta_{\tau'_2 \tau_2} X_{\tau'_1 \tau'_3 \tau_1 \tau_3}+\delta_{\tau'_3 \tau_3} X_{\tau'_1 \tau'_2 \tau_1 \tau_2}.
\eeqn{bigXi}

\subsection{3N force containing the $(\ve{\sigma}_i \cdot \ve{\sigma}_j)$ terms.}

Here we will derive an expression for matrix element of spin-dependent hypercentral 3N force 
\beq
\fl
W_{123}^{(\sigma)} \equiv \sum_{i\neq k<j \neq k} (\ve{\sigma}_i \cdot \ve{\sigma}_j) W_{k,ij}^{(\sigma)} = \left[
(\ve{\sigma}_1 \cdot \ve{\sigma}_2)+(\ve{\sigma}_1 \cdot \ve{\sigma}_3)+(\ve{\sigma}_2 \cdot \ve{\sigma}_3)\right] W_{123}^{(\sigma)}(\rho_{123}).
\eeqn{Wsigma}
We will start with contribution from the $(\ve{\sigma}_1 \cdot \ve{\sigma}_2) W_{123}(\rho_{123})$ term. We will use Eq. (\ref{tautau}), in which $\tau$ is replaced by $\sigma$, then Eq. (\ref{summ3}). We also need to perform two other sums over angular momentum projections, 
\beq
\fl
\sum_
{ \scriptsize{ \begin{array} {ll} {m_1 \sigma_1  }\\{ m'_1 \sigma'_1}\end{array}}}
(-)^{m'_1+\sigma_1+\dem} (l'_1 \, {\rm -}m'_1 l_1 m_1 | L_1 M_1) \, (l'_1m'_1 \dem \sigma'_1 | j'_1 m'_{j_1})
(l_1m_1 \dem \sigma_1 | j_1 m_{j_1})
( \dem \sigma'_1 \,\,\dem \,  {\rm -}\sigma_1 \mid 1 s)
\eol  
= \sum_{J_1 M_{J_1}} (-)^{l_1+j_1-m_{j_1}+J_1+1} \,\hat{j}_1 \hat{j}'_1\hat{L}_1 \hat{J}_1 (j'_1 \,  {\rm -}m'_{j_1} j_1 m_{j_1} | J_1 M_{J_1})
\eol\,\,\,\,\,\,\,\,\,\,\,\,\,\,\,\, \times
(J_1 M_{J_1} L_1\,  {\rm -}M_1 | 1 \,  {\rm -}s) \left\{ \begin{array}{lll} {l_1} &{ \frac{1}{2} }&{j_1}\\
{l'_1 }&{\frac{1}{2}}&{j'_1 }
\\{L_1}&{1}&{J_1}\end{array}\right\}
\eeqn{sum1}
and
\beq
\fl
\sum_{M_1M_2s} (-)^{L_1+M_1+L_2+M_2+s} (J_1 M_{J_1} L_1\,  {\rm -}M_1 | \lambda s) (J_2 M_{J_2} L_2\,  {\rm -}M_2 | \lambda \,  {\rm -}s)
  (L_1M_1L_2M_2| L_{3}M_{3})
\eol
= (-)^{\lambda +L_1+J_1+L_2+J_2} \, \hat{\lambda}^2 (J_1 M_{J_1} J_2 M_{J_2}|  L_{3}M_{3}) W(J_1 \lambda L_3 L_2;L_1J_2).
\eeqn{sum2}
This results in
\beq
\fl
\la \varphi_{\alpha'_1}(\ve{r}_1) \varphi_{\alpha'_2}(\ve{r}_2) e^{i \ve{k}\ve{r}_3} \chi_{\tau'_3}
\mid 
(\ve{\sigma}_1 \cdot \ve{\sigma}_2) W_{123}
\mid
\varphi_{\alpha_1}(\ve{r}_1)\varphi_{\alpha_2}(\ve{r}_2)\varphi_{\alpha_3}(\ve{r}_3)\ra =
\eol
= \delta_{\tau'_1\tau_1}\delta_{\tau'_2\tau_2}\delta_{\tau'_3\tau_3} \sum_{n'_3 l'_3j'_3 m'_{j_3}} \varphi_{n'_3 l'_3 j'_3m'_{j_3}} (\theta \ve{k})
\sum_{L_1L_2L_3} W^{(\sigma)}_{\gamma_1\gamma_2\gamma_3}
S_{\eta_1 \eta_2}^{(\eta_3)},
\eeqn{x3}
where $\eta_i = \{ l_ij_im_{j_i}l'_ij'_im'_{j_i}L_i\}$ and
\beq
\fl
S_{\eta_1 \eta_2}^{(\eta_3)} = 
 (-)^{l_1+j_1-m'_{j_1}+l_2+j_2-m'_{j_2}+l_3+j_3-m'_{j_3}}  (-)^{L_3+M_3}\,\, 6\hat{j}_1\hat{j}'_1\hat{j}_2\hat{j}'_2\hat{j}_3\hat{j}'_3 \hat{L}_1 \hat{L}_2 \hat{L}_3^{-1}
\eol 
\fl\times
\sum_{J_1 J_2} \hat{J}_1 \hat{J}_2 
(j'_1 -m'_{j_1} j_1 m_{j_1} \mid J_1M_{J_1})(j'_2 -m'_{j_2} j_2 m_{j_2} \mid J_2M_{J_2})
(j'_3 -m'_{j_3} j_3 m_{j_3} \mid L_3M_3)
\eol 
\fl\times
\,\,\,\,\,\,\,\,\,\,\,\,\,\,\,(J_1 M_{J_1} J_2 M_{J_2} \mid L_3 -M_3) 
W(L_3l_3 j'_3 \dem ; l'_3 j_3) W(J_1 1 L_3 L_2 1; L_1 J_2) \eol \times
\left\{ \begin{array}{lll} {l_1} &{ \frac{1}{2} }&{j_1}\\
{l'_1 }&{\frac{1}{2}}&{j'_1 }
\\{L_1}&{1}&{J_1}\end{array}\right\}
\left\{ \begin{array}{lll} {l_2} &{ \frac{1}{2} }&{j_2}\\
{l'_2 }&{\frac{1}{2}}&{j'_2 }
\\{L_2}&{1}&{J_2}\end{array}\right\}. 
\eeqn{x4}
To obtain this result we used $\lambda = 1$,   $m_{j_1}+m_{j_2}+m_{j_3} = m'_{j_1}+m'_{j_2}+m'_{j_3}  $ and that for a system of fermions relation $(-)^{2m_{j}+1} = 1$ is valid for any angular momentum  projection $m_j$. In the limit of $\lambda = 0$ we recover results from previous subsection.

In a similar fashion we can get  expressions for contributions from $(\ve{\sigma}_1\cdot\ve{\sigma}_3)$ and $(\ve{\sigma}_2\cdot\ve{\sigma}_3)$. Then the final expression for contribution from $W^{(\sigma)}_{123}$ is
\beq
\fl
\la \varphi_{\alpha'_1}(\ve{r}_1) \varphi_{\alpha'_2}(\ve{r}_2) e^{i \ve{k}\ve{r}_3} \chi_{\tau'_3}
\mid 
W^{(\sigma)}_{123}
\mid
\varphi_{\alpha_1}(\ve{r}_1)\varphi_{\alpha_2}(\ve{r}_2)\varphi_{\alpha_3}(\ve{r}_3)\ra =
\eol
\fl 
= \delta_{\tau'_1\tau_1}\delta_{\tau'_2\tau_2}\delta_{\tau'_3\tau_3} \sum_{n'_3 l'_3j'_3 m'_{j_3}} \varphi_{n'_3 l'_3 j'_3m'_{j_3}} (\theta \ve{k})
\sum_{L_1L_2L_3} W^{(\sigma)}_{\gamma_1\gamma_2\gamma_3} \left(
S_{\eta_1 \eta_2}^{(\eta_3)} + S_{\eta_1 \eta_3}^{(\eta_2)} + S_{\eta_2 \eta_3}^{(\eta_1)} \right),
\eeqn{x5}

\subsection{3N force containing the $(\ve{\sigma}_i \cdot \ve{\sigma}_j)(\ve{\tau}_i \cdot \ve{\tau}_j)$ terms.}
Finally, using results from previous subsections it is easy to obtain expression for the matrix element of the spin- and isospin-dependent 3N force
\beq
\fl
W_{123}^{(\sigma\tau)} \equiv
\sum_{i\neq k<j \neq k} (\ve{\sigma}_i \cdot \ve{\sigma}_j)(\ve{\tau}_i \cdot \ve{\tau}_j) W_{k,ij}^{(\sigma\tau)} 
\eol 
\fl \,\,\,\,\,\,\,\,\,\,\,\,\,\,\,\,\, 
= \left[
(\ve{\sigma}_1 \cdot \ve{\sigma}_2)(\ve{\tau}_1 \cdot \ve{\tau}_2)+(\ve{\sigma}_1 \cdot \ve{\sigma}_3)(\ve{\tau}_1 \cdot \ve{\tau}_3)+(\ve{\sigma}_2 \cdot \ve{\sigma}_3)(\ve{\tau}_2 \cdot \ve{\tau}_3)\right] W_{123}^{(\sigma\tau)}(\rho_{123}).
\eeqn{Wsigmatau}
This expression is
\beq
\fl
\la \varphi_{\alpha'_1}(\ve{r}_1) \varphi_{\alpha'_2}(\ve{r}_2) e^{i \ve{k}\ve{r}_3} \chi_{\tau'_3}
\mid 
W_{123}^{(\sigma\tau)}
\mid
\varphi_{\alpha_1}(\ve{r}_1)\varphi_{\alpha_2}(\ve{r}_2)\varphi_{\alpha_3}(\ve{r}_3)\ra =
\eol
=  \sum_{n'_3 l'_3j'_3 m'_{j_3}} \varphi_{n'_3 l'_3 j'_3m'_{j_3}} (\theta \ve{k})
\sum_{L_1L_2L_3} W_{\gamma_1\gamma_2\gamma_3}  
\eol\times
\left( \delta_{\tau'_3\tau_3}X_{\tau'_1\tau'_2 \tau_1 \tau_2}
S_{\eta_1 \eta_2}^{(\eta_3)} + \delta_{\tau'_2\tau_2}X_{\tau'_1\tau'_3 \tau_1 \tau_3}S_{\eta_1 \eta_3}^{(\eta_2)} + \delta_{\tau'_1\tau_1}X_{\tau'_2\tau'_3 \tau_2 \tau_3}S_{\eta_2 \eta_3}^{(\eta_1)} \right).
\eeqn{MEst}

%\subsection{Limit $\rho_0 \rightarrow \infty$}
% Limit to $\rho_0 \rightarrow \infty$.
% \beq
%\la \Phi_{A-1} | W_0| \Phi_A \ra \rightarrow %\frac{(A-1)(A-2)}{2}  \la \Phi_{A-1} | \Phi_A \ra
%\eeqn{}
%\\
%$\la \Phi_{A-1} | W_{\tau}| \Phi_A \ra \rightarrow \left[(A-2)(4T_A(T_A+1)-3)+2T_{A-1}(T_{A-1}+1) -\frac{3}{2}(A-1)\right]  \la \Phi_{A-1} | \Phi_A \ra$


\section*{References}
\begin{thebibliography}{}
\bibitem{TJPPNP} Timofeyuk N K and Johnson R C   2020 {\em Prog. Part. Nucl. Phys.} 111 103738

\bibitem{Tim14} Timofeyuk N K 2014 {\em J.Phys.G: Nucl.Part.Phys.} 41 094008 


\bibitem{Nav11} Navratil P, Roth R and Quaglioni S 2011 \PL {\bf B} 704 379
\bibitem{Cal16}  Calci A \etal   2016 \PR {\bf C} {117} {42501} %{2016}
\bibitem{Pin65} Pinkston W T and Satchler G R 1965 \NP 72, 641
\bibitem{Phi68}  Philpott R J,  Pinkston W T and  Satchler G R, 1968 \NP {\bf A} {119} {241} %{1968}
\bibitem{Ban85} Bang J M, Gareev F G, Pinkston W T and Vaagen J S, 1985,  Phys Rep 125, 253
\bibitem{Tim98}  Timofeyuk N K 1998 \NP {\bf A} {19} {632} %{1998}
\bibitem{Tim09} Timofeyuk N K 2009 \PRL {103} {242501} %{2009}

\bibitem{Tim10} Timofeyuk N K  2010 \PR {\bf C} {81} {064306} %{2010}
\bibitem{Tim11} Timofeyuk N K 2011 \PR {\bf C} {84} {054313} 

\bibitem{Tim13}  Timofeyuk N K 2013 \PR {\bf C} {88} {044315} 

\bibitem{Mac60}  MacFarlaine M H  and  French J B 1960 {\RMP}  {32} {567} %{1960} 

\bibitem{Gom18}  G\'omez-Ramos M, Moro A M 2018 \PL {\bf B}  785, 511

\bibitem{Vaq20}  Vaquero V et al. 2020 \PRL
 124 022501
 
 \bibitem{Fel98}  Feldemeier H,   Neff T,   Roth R,   Schnack J 1998 \NP {\bf A} 632 61 
 
 \bibitem{Fur12}   Furnstahl R J 2012 \NP {\bf B}  Proc. Suppl.
 228  139 

\bibitem{M3Y}
 Bertsch G, Borysowicz J, McManus H and Love W G 1977 \NP {\bf A}  284 399 
\bibitem{Elliot}  Elliott J P, Jackson A D, Mavromatis H A, Sanderson E A,
and Singh B  1968 \NP {\bf A} 121 241 

\bibitem{Blo68}   Blomqvist J and   Molinari A 1968 Nucl. Phys. A 106, 545 
\bibitem{Kar10}  Karataglidis S, Henninger K R, Richter W A, Amos K, 2010 \NP {\bf A} 848  110 
\bibitem{Kar02}  Karataglidis S, Amos K, Brown B A, Deb P K 2002 \PR {\bf C} 65 044306  
\bibitem{Kar01}  Karataglidis S and   Chadwick M B 2001 \PR {\bf C} 64, 064601  
\bibitem{Dor98}  Dortmans P J,  Amos K,  Karataglidis S and Raynal J 1998 \PR {\bf C} 58  2249  
\bibitem{Dor97} Dortmans P J, Amos K, Karataglidis S 1997  J.Phys.(London) G23 183 


%\bibitem{Atk18}  Atkinson M C,  Blok H P,  Lapik\'as L,  Charity R J and  Dickhoff W H  2018 \PR {\bf C} 98  044627  

\bibitem{Kra01}   Kramer G J,   Blok H P  and   Lapik\'as L 2001 \NP {\bf A}   679 267 

\bibitem{Art03}   Artemov S V,   Zaparov E A, and   Nie G K 2003 Bull. Rus. Acad.
Sci., Phys. 67  1741 

\bibitem{Muk08}  Mukhamedzhanov A M et al. 2008 \PR {\bf C} 78  015804  

\bibitem{Leu94}   Leuschner M et al. 1994 \PR {\bf C} 49 955 


\bibitem{Moshinsky}  Moshinsky M 1959 \NP  13 104 

\bibitem{Smirnov} Smirnov Yu F 1962 \NP   39  346  
\bibitem{Trlifaj} Trlifaj L 1972 \PR {\bf C} 5 1534 

\end{thebibliography}
\end{document}